\documentclass[journal]{IEEEtran}
\IEEEoverridecommandlockouts
\usepackage{lettrine}
\usepackage{cite}
\usepackage{amsmath,amssymb,amsfonts}
\usepackage{algorithm}
\usepackage{algorithmic}
\usepackage{bm}
\usepackage{graphicx}
\usepackage{subfigure}
\usepackage{tabularx}
\usepackage{textcomp}
\usepackage{xcolor}
\usepackage{stfloats}
\usepackage{multirow}
\def\BibTeX{{\rm B\kern-.05em{\sc i\kern-.025em b}\kern-.08em
		T\kern-.1667em\lower.7ex\hbox{E}\kern-.125emX}}

\floatname{algorithm}{\bf{Algorithm}}

\begin{document}

\title{Joint Active and Passive Beamforming for IRS-Aided Wireless Energy Transfer Network Exploiting One-Bit Feedback}

\author{Taotao~Ji, Meng~Hua, Chunguo~Li,~\IEEEmembership{Senior~Member,~IEEE,} Yongming~Huang,~\IEEEmembership{Senior~Member,~IEEE,} and~Luxi~Yang,~\IEEEmembership{Senior~Member,~IEEE}%
	\IEEEcompsocitemizethanks{\IEEEcompsocthanksitem An earlier version of this paper was presented in part at the IEEE GLOBECOM 2023.
    \IEEEcompsocthanksitem Taotao Ji, Chunguo Li, Yongming Huang, and Luxi Yang are with the School of Information Science and Engineering, the National Mobile Communications Research Laboratory, and the Frontiers Science Center for Mobile Information Communication and Security, Southeast University, Nanjing 210096, China, and also with the Pervasive Communications Center, Purple Mountain Laboratories, Nanjing 211111, China (e-mail: jitaotao@seu.edu.cn; chunguoli@seu.edu.cn; huangym@seu.edu.cn; lxyang@seu.edu.cn).
	\IEEEcompsocthanksitem Meng Hua is with the Department of Electrical and Electronic Engineering, Imperial College London, London SW7 2AZ, UK (e-mail: m.hua@imperial.ac.uk).
}}
\maketitle
\begin{abstract}
To reap the active and passive beamforming gain in an intelligent reflecting surface (IRS)-aided wireless network, a typical way is to first acquire the channel state information (CSI) relying on the pilot signal, and then perform the joint beamforming design. However, it is a great challenge when the receiver can neither send pilot signals nor have complex signal processing capabilities due to its hardware limitation. To tackle this problem, we study in this paper an IRS-aided wireless energy transfer (WET) network and propose two joint beamforming design methods, namely, the channel-estimation-based method and the distributed-beamforming-based method, that require only one-bit feedback from the energy receiver (ER) to the energy transmitter (ET).
Specifically, for the channel-estimation-based method, according to the feedback information, the ET is able to infer the cascaded ET-IRS-ER channel by continually adjusting its transmit beamformer while applying the analytic center cutting plane method (ACCPM). Then, based on the estimated cascaded CSI, the joint beamforming design can be performed by using the existing optimization techniques. While for the distributed-beamforming-based method, we first apply the distributed beamforming algorithm to optimize the IRS reflection coefficients, which is theoretically proven to converge to a local optimum almost surely. Then, the optimal ET’s transmit covariance matrix is obtained based on the effective ET-ER channel learned by applying the ACCPM only once. Numerical results demonstrate the effectiveness of our proposed one-bit-feedback-based joint beamforming design schemes while greatly reducing the requirement on the hardware complexity of the ER. In particular, the high accuracy of our IRS-involved cascaded channel estimation method exploiting one-bit feedback is also validated.
\end{abstract}

\begin{IEEEkeywords}
Intelligent reflecting surface (IRS), one-bit feedback, analytic center cutting plane method (ACCPM), distributed beamforming, channel estimation, wireless energy transfer (WET).
\end{IEEEkeywords}

\section{Introduction}
\IEEEPARstart{T} {o achieve} the vision that descripted as ``global coverage, all spectra, full applications, all senses, all digital, and strong security" for the sixth-generation (6G) wireless communication systems \cite{10054381_Wang}, the traditional thinking that wireless communication technology needs to adapt to uncontrollable wireless channels should be broken.
Intelligent reflecting surface (IRS), with the ability to smartly reshape the electromagnetic propagation environment in real time to enhance wireless system performance, has been proposed as a promising new paradigm for the future 6G wireless networks \cite{9903378_Chen,9475160_Pan,9530717_Basharat,9326394_Wu,9570143_Mu}. Typically, IRSs are implemented by a large number of low-cost passive elements (e.g., positive-intrinsic-negative (PIN) diodes, field-effect transistors (FETs), or micro-electromechanical system (MEMS) switches \cite{9326394_Wu}), and do not rely on active hardware components such as radio frequency (RF) chains. 
By finely adjusting the phase shifts (PSs) of the passive IRS reflecting elements, the incident signals can be dynamically shaped to meet diverse system requirements, including the spectral and energy efficiency improvements \cite{10136805_Ji,9727083_Wu,9690055_Chen,9913311_Hua}, communication coverage expansion \cite{9790792_Ma,9980412_Shi,9201413_Zeng}, physical layer security enhancement \cite{9146177_Wang,9769929_Qiao}, etc. In general, the introduction of IRS brings about a new design degree of freedom (DoF) for improving the quality of channels to existing wireless systems without modifying the current physical layer standardization. Moreover, IRS has attractive characteristics such as low profile, light-weight, and conformal geometry, thus allowing its highly flexible deployment. All of the above significant advantages make IRS a promising research direction for both industry and academia.

Typically, to facilitate resource allocation and channel reconfiguration in an IRS-aided wireless network, a prerequisite is the acquisition of IRS-involved cascaded channel state information (CSI).
Accordingly, the expected system performance gains brought by an IRS depend on accurate CSI estimation, which is a challenge since IRS is a passive component that can neither send nor receive pilot symbols, and its elements number is typically very large \cite{9771077_Swindlehurst}.
Take the downlink wireless transmission as an example, a widely adopted channel estimation approach is that for the systems operating in time-division duplex (TDD) protocol, the transmitter acquires the IRS-involved cascaded CSI by estimating the reverse link based on the training signals sent by the receiver \cite{9053695_Jensen,9505267_Liu,10014657_Zhang}. However, this method critically depends on the degree of reciprocity between the forward and reverse links. Moreover, to improve the channel estimation accuracy, the receiver needs to send pilot symbols with sufficient power to effectively mitigate the impact of noise, which is unbearable for energy-constrained receiver, such as mobile phones in cellular network. 
Alternatively, another commonly adopted method to acquire the IRS-related cascaded CSI at the transmitter is to send the pilot signals from transmitter to receiver, through which the receiver can estimate the CSI and then send it back to the transmitter via a feedback channel \cite{9103231_Wang,9361077_de}. Although this method applies to both TDD and frequency-division duplex (FDD) adopted systems, it requires complex signal processing capabilities at the receiver for channel estimation, thus increasing its hardware cost. Responding to this, the beam training strategy is a desired method that facilitates both efficient channel estimation and beamforming design while reducing the hardware cost at the receiver \cite{10057262_Wang,9410435_Wang,9325920_Ning}. However, it mainly takes advantage of the sparsity of millimeterwave (mmWave) or terahertz (THz) channels and can only realize the mainlobe alignment, thus it is not suitable for low-frequency wireless systems. The drawbacks of the existing channel estimation methods described above motivate us to investigate novel channel learning and beamforming approaches for IRS-aided wireless systems while taking the hardware limitation at the receiver into consideration in our work.

In this paper, we study an IRS-aided wireless energy transfer (WET) network, where an energy transmitter (ET) sends wireless energy to an energy receiver (ER) via transmit energy beamforming with the help of an IRS. It is assumed that the ER is unable to assist the ET to perform explicit channel estimation due to its hardware limitation\footnote{As an example, it seems impossible for an ER with the architecture similar to that in \cite{6623062_Zhou} to incorporate baseband signal processing for channel estimation.}, while it can periodically send one-bit feedback information, i.e., `0' or `1', to the ET over the dedicated feedback channel to help improve system performance. Based on this system setup, we propose two novel joint active ET and passive IRS beamforming design methods, namely, the channel-estimation-based method and the distributed-beamforming-based method, to boost the system performance.

For the channel-estimation-based method, we first estimate the cascaded ET-IRS-ER channel by continuously adjusting the transmit beamformer of the ET according to the feedback information from the ER each time. Each feedback bit indicates whether the energy harvested at the ER in the current interval is increasing or decreasing compared to the previous interval, which we assume can be accurately measured. It is worth noting that an optimization technique named analytic center cutting plane method (ACCPM) is applied during the channel learning phase. Then, in the following wireless energy transmission phase, based on the estimated channel, the transmit covariance matrix of the ET and the PSs of the IRS can be jointly optimized to maximize the amount of energy harvested at the ER by applying the existing optimization techniques.
It should be mentioned that although the channel learning algorithms that require one-bit feedback were respectively proposed in \cite{6884811_Xu} and \cite{7117443_Gopalakrishnan}, IRS was not applied in their studied systems, which as a passive component that greatly increases the difficulty of channel estimation. 

For the distributed-beamforming-based method, we jointly optimize the ET's transmit covariance matrix and the IRS reflection coefficients without explicitly estimating the cascaded ET-IRS-ER channel. Specifically, we first optimize the IRS reflection coefficients by applying the distributed beamforming algorithm, where each IRS element/group adjusts its PS value randomly at each iteration, and the ER gives one-bit of feedback indicating whether the amount of harvested energy is larger or smaller than that in previous iteration. If it is larger, all the IRS elements/groups will keep their latest PS perturbations; otherwise they all undo the PS perturbation. The above procedure is repeated until the convergence is reached. In addition, we theoretically show that for trivial perturbation distribution, with any given initial value, the PSs of IRS will converge to a local optimum with probability 1. Our feedback algorithm for optimizing the PSs of IRS elements can be seen as an extension of the distributed transmit beamforming algorithm proposed in \cite{5361473_Mudumbai} since the optimization variables are coupled in the objective function, which makes the proof of convergence more difficult. Then, based on the optimized IRS reflection coefficients, the effective ET-ER channel can be asymptotically obtained by using the ACCPM, and the optimal ET's transmit covariance matrix can be therewith derived in closed form solution. It is worth noting that the above joint active and passive beamforming optimization procedures only need to be iterated once. 

Finally, we perform extensive simulations to demonstrate the effectiveness of our proposed one-bit-feedback-based joint beamforming design methods, which significantly reduce the requirement for the hardware complexity of the ER simultaneously. In particular, it is also shown that our proposed channel-estimation-based method is able to estimate the cascaded ET-IRS-ER channel with high precision.

The rest of this paper is organized as follows. Section II introduces the system model and the problem formulation for joint active and passive beamforming design in IRS-aided WET network. In Section III and IV, we propose two efficient joint beamforming design methods, namely, the channel-estimation-based method and the distributed-beamforming-based method, based on one-bit feedback. Section V presents numerical results to evaluate the performance of the proposed joint beamforming design schemes. Finally, Section VI concludes the paper.

\emph{Notations:}
$ \bf{I} $, $ \bf{0} $, and $ \bf{1} $ denote an identity matrix, an all-zero
matrix, and an all-one matrix, respectively. $ \mathbb{C}^{x \times y} $ denotes the space of $ x \times y $ complex-valued matrices.
For a complex number $ x $, $ \Re \left\{ x \right\} $, $ \Im \left\{ x \right\} $, $ \left| x \right| $, $ \arg \left( x \right) $ denote its real part, imaginary part, modulus, and phase, respectively.
For a complex-valued vector $ \bf{x} $, $ \left\| {\bf{x}} \right\| $ denotes its Euclidean norm.
For any general matrix $ \bf{X} $, $ {\left\| {\bf{X}} \right\|_F} $, $ {{\bf{X}}^H} $, $ {\rm{rank}}\left( {\bf{X}} \right) $, and $ {\bf{X}}\left( {i,j} \right) $ denote its Frobenius norm, conjugate transpose, rank, and $ \left( {i,j} \right) $th element, respectively.
For a square matrix $ \bf{A} $, $ \det \left( {\bf{A}} \right) $, $ {\rm{tr}}\left( {\bf{A}} \right) $ and $ {{\bf{A}}^{ - 1}} $ denote its determinant, trace, and inverse, respectively, and $ {\bf{A}} \succeq {\bf{0}} $ means that $ \bf{A} $ is positive semi-definite.
$ {\rm{diag}}\left\{ {{a_1}, \cdots ,{a_n}} \right\} $ denotes a square diagonal matrix with $ {{a_1}, \cdots ,{a_n}} $ on the diagonal. $ \mathbb{E}\left\{  \cdot  \right\} $ and $ \Pr \left(  \cdot  \right) $ denote the statistical expectation and probability function, respectively. The Kronecker product and Khatri-Rao product are denoted by $  \otimes  $ and $  \odot  $, respectively. 
The distribution of a circularly symmetric complex Gaussian (CSCG) random vector with mean vector $ \bf{x} $ and covariance matrix $ {\bf{\Sigma }} $ is
denoted by $ {\cal C}{\cal N}\left( {{\bf{x}},{\bf{\Sigma }}} \right) $, and $  \sim  $ stands for ``distributed as”.

\begin{figure}[t]
	\centerline{\includegraphics[width=0.38\textwidth]{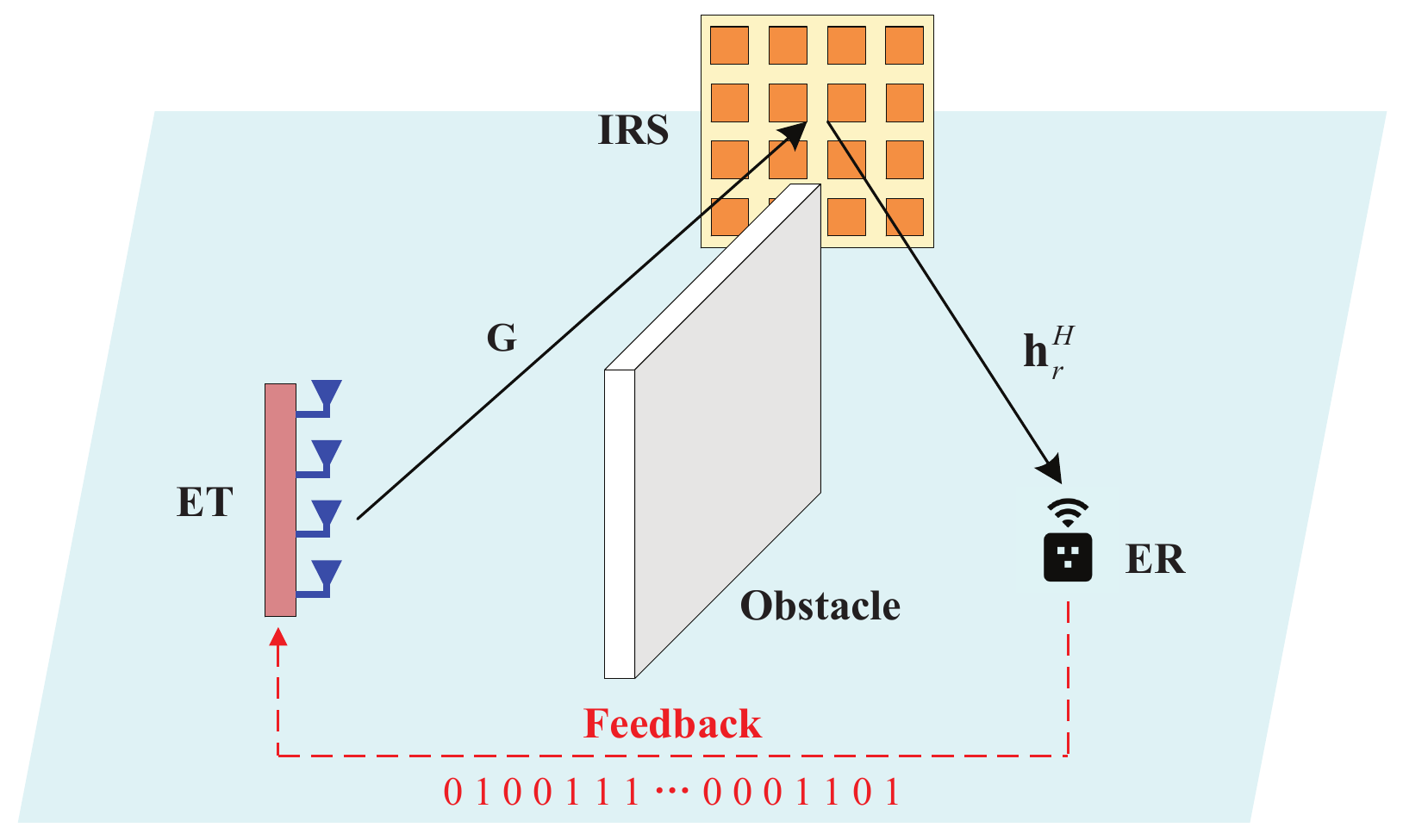}}
	\caption{An indoor IRS-aided WET system where the ER feeds back one-bit of information to the ET periodically to assist the joint beamforming design for the system.}
	\label{fig_system_model}
\end{figure}

\section{System Model And Problem Formulation}

As shown in Fig. \ref{fig_system_model}, we consider a typical indoor IRS-assisted WET network, where an IRS consisting of $ N $ passive reflecting elements is deployed to assist in the WET from an ET equipped with $ M_t \ge 1 $ antennas to a single-antenna ER. In particular, we consider a special case where the direct ET-to-ER transmission is blocked due to the unfavorable propagation conditions.
Due to the hardware limitation at the ER, it can neither send pilot symbols nor perform complex signal processing operations, which makes the studied system fail to complete explicit channel estimation. However, the ER can periodically feed back one-bit of information to the ET based on the communication protocol to help the system improve its WET efficiency.
The baseband equivalent channels from the ET to the IRS, from the IRS to the ER are denoted by $ {\bf{G}} \in {\mathbb{C}^{N \times {M_t}}} $ and $ {\bf{h}}_r^H \in {\mathbb{C}^{1 \times N}} $, respectively. Since the ET, IRS, and ER are all placed in fixed positions, we assume that both the ET-IRS channel, $ \bf{G} $, and the IRS-ER channel, $ {\bf{h}}_r^H $, are quasi-static. Denote the duration of each channel coherence interval (CCI) as $ T_c $, it is assumed to be sufficiently long for our proposed joint active and passive beamforming design schemes exploiting only one-bit feedback. We also assume that the ET can send $ N_d \le M_t $ energy beams, where the value of $ N_d $ can be changed as needed. The $ m $th beamforming vector and its corresponding energy-carried signal are denoted by $ {{\bf{w}}_m} \in {\mathbb{C}^{{M_t} \times 1}} $ and $ {s_m} \in \mathbb{C} $, respectively, with $ m \in {{\cal N}_d} $ and $ {{\cal N}_d} \buildrel \Delta \over = \left\{ {1,2, \cdots ,{N_d}} \right\} $. $ \left\{ {{s_m}} \right\}_{m = 1}^{{N_d}} $ can be assumed to be independent random variables from any arbitrary distribution with zero mean and unit variance, i.e., $ \mathbb{E}\big\{ {{{\left| {{s_m}} \right|}^2}} \big\} = 1,\forall m \in {{\cal N}_d} $, due to the fact that they do not carry any information. Thus, the transmitted signal at the ET is give by $ {\bf{x}} = \sum\nolimits_{m = 1}^{{N_d}} {{{\bf{w}}_m}{s_m}} $. Denote the set of reflecting elements of the IRS as $ {\cal N}  \buildrel \Delta \over =  \left\{ {1,2, \cdots N} \right\} $. The IRS reflection coefficient matrix can be modeled as $ {\bf{\Theta }} = {\rm{diag}}\left\{ {{e^{j{\theta _1}}},{e^{j{\theta _2}}}, \cdots ,{e^{j{\theta _N}}}} \right\} $, where $ {\theta _n} \in \left[ {0,2\pi } \right),\forall n \in {\cal N} $, represents the PS corresponding to the $ n $th IRS reflecting element. Accordingly, the ER's received signal at time $ t $ is given by
\begin{align}
y\left( t \right) = {\bf{h}}_r^H{\bf{\Theta }}\left( t \right){\bf{Gx}}\left( t \right) + n\left( t \right),
\end{align}
where $ n \left( t \right) \sim {\cal C}{\cal N}\left( {0,\sigma _n^2} \right) $ denotes the additive white Gaussian noise received by the ER at time $ t $. Let $ {\bf{v}} = {\left[ {{v_1},{v_2}, \cdots {v_N}} \right]^H} $, where $ {v_n} = {e^{j{\theta _n}}} $, thus we have the following unit-modulus constraints: $ \left| {{\bf{v}}\left( n \right)} \right| = 1,\forall n \in \mathcal{N} $.
By applying the change of variables, we have
\begin{align}
y\left( t \right) &= {{\bf{v}}^H}\left( t \right){\rm{diag}}\left( {{\bf{h}}_r^H} \right){\bf{Gx}}\left( t \right) + n\left( t \right) \notag\\
&= {{\bf{v}}^H}\left( t \right){{\bf{H}}_c}{\bf{x}}\left( t \right) + n\left( t \right),
\end{align}
where $ {{\bf{H}}_c} = {\rm{diag}}\left( {{\bf{h}}_r^H} \right){\bf{G}} \in {\mathbb{C}^{N \times {M_t}}} $ is defined as the cascaded ET-IRS-ER channel. The energy harvested from noise is assumed to be negligible\footnote{Or the ER can measure the amount of its harvested energy averaged over a large number of symbols to minimize the effect of noise.}, thus the amount of the energy harvested at the
ER is given by
\begin{align}
{E_h}\left( t \right) = \eta{\rm{tr}}\left( {{{\bf{v}}^H}\left( t \right){{\bf{H}}_c}{\bf{Q}}\left( t \right){\bf{H}}_c^H{\bf{v}}\left( t \right)} \right),
\end{align}
where $ {\bf{Q}} = \mathbb{E}\big\{ {{\bf{x}}{{\bf{x}}^H}} \big\} = \sum\nolimits_{m = 1}^{{N_d}} {{{\bf{w}}_m}{\bf{w}}_m^H} \succeq {\bf{0}} $ is the transmit covariance matrix at the ET, and $ 0 < \eta  \le 1 $ represents the energy harvesting efficiency. Since the ET has a maximum transmit power budget denoted by $ P_b $, then we have $ {\rm{tr}}\left( {\bf{Q}} \right) \le {P_b} $. Our objective is to maximize the amount of energy harvested at the ER without explicitly knowing the cascaded ET-IRS-ER channel $ {{\bf{H}}_c} $, by jointly optimizing the transmit covariance matrix $ {\bf{Q}} $ at the ET and the PS vector $ \bf{v} $ at the IRS, while subject to the total transmit power constraint at the ET. As a result, we can formulate the problem as
\begin{subequations}\label{P1}
\begin{align}
\mathop {\max }\limits_{{\bf{Q}},{\bf{v}}} \ &\eta {\rm{tr}}\big( {{{\bf{v}}^H}{{\bf{H}}_c}{\bf{QH}}_c^H{\bf{v}}} \big)\\
{\rm{s}}{\rm{.t}}{\rm{. }}&\left| {{\bf{v}}\left( n \right)} \right| = 1,\forall n \in {\cal N},\\
&{\rm{tr}}\left( {\bf{Q}} \right) \le {P_b},\\
&{\bf{Q}} \succeq {\bf{0}}.
\end{align}
\end{subequations}
The key challenge for solving problem (\ref{P1}) lies in that we cannot explicitly obtain the required cascaded ET-IRS-ER channel $ {{{\bf{H}}_c}} $, and there is only periodic one-bit feedback from the ER to the ET to be utilized to help adjust $ \bf{Q} $ and $ \bf{v} $. To tackle this intractability, in the next two sections we propose two novel joint active and passive beamforming design methods, namely, the channel-estimation-based method and the distributed-beamforming-based method, for our studied IRS-aided WET system by exploiting the ER's one-bit feedback.

\section{Channel-Estimation-Based Method}
In this section, we propose a novel channel-estimation-based joint active and passive beamforming design scheme for the IRS-aided WET network exploiting only one-bit feedback from the ER to the ET. Specifically, we first estimate a scaled version of the cascaded ET-IRS-ER channel by continuously adjusting $ {\bf{Q}} $ and $ {\bf{v}} $ according to the periodic one-bit feedback information from the ER to ET. Then, the optimal $ {\bf{Q}} $ and $ {\bf{v}} $ can be obtained by applying the existing optimization methods based on the estimated channel.

\subsection{IRS Elements Grouping}
It can be readily seen that the dimension of the cascaded channel $ {{{\bf{H}}_c}} $ grows linearly with the number of IRS elements, $ N $, which can be huge in practice. For this, we adopt an IRS elements grouping method to reduce the channel learning overhead\footnote{The rationality lies in that the IRS elements are usually tightly packed, and thus the channels for adjacent elements are highly correlated \cite{9039554_Yang}.}.
The IRS elements in the same group are assigned a common reflection coefficient. Denote by $ J $ the number of groups, we have
\begin{align}
{{\bf{v}}^H}{{\bf{H}}_c} = \left( {{{{\bf{\bar v}}}^H} \odot \left[ {{{\bf{1}}_{1 \times {K_1}}},{{\bf{1}}_{1 \times {K_2}}}, \cdots ,{{\bf{1}}_{1 \times {K_J}}}} \right]} \right){{\bf{H}}_c},
\end{align}
where $  \odot  $ stands for the Khatri-Rao product, $ {\bf{\bar v}} = {\left[ {{{\bar v}_1},{{\bar v}_2}, \cdots ,{{\bar v}_J}} \right]^T} \in {\mathbb{C}^{J \times 1}} $ represents the IRS group reflection coefficients, with $ {{\bar v}_j} $ denoting the common reflection coefficient for the $ j $th group, and $ {K_j},\forall j \in {\cal J} $, is the number of the IRS elements in the $ j $th group, with $ {\cal J}  \buildrel \Delta \over =  \left\{ {1,2, \cdots ,J} \right\} $. Obviously, we have $ \sum\nolimits_{j = 1}^J {{K_j}}  = N $. In our work, we assume equal size (number of IRS elements) of each group, which is given by\footnote{We assume $ K $ is an integer without loss of generality.} $ K = N/J $. Therefore, we have
\begin{align}
{{\bf{v}}^H}{{\bf{H}}_c} &= \left( {{{{\bf{\bar v}}}^H} \otimes {{\bf{1}}_{1 \times K}}} \right){{\bf{H}}_c} \notag\\
&= \sum\limits_{i = 1}^N {{{\left( {{{{\bf{\bar v}}}^H} \otimes {{\bf{1}}_{1 \times K}}} \right)}_i}{\bf{h}}_{c,i}^H}   \notag \\
&= \sum\limits_{i = 1}^J {{\bf{\bar v}}_i^ * \sum\limits_{j = 1}^K {{\bf{h}}_{c,j + \left( {i - 1} \right)K}^H} }  \notag \\
&= {{{\bf{\bar v}}}^H}{{{\bf{\bar H}}}_c},
\end{align}
where $ {{\bf{h}}_{c,i}^H} $ is the $ i $th row of $ {{\bf{H}}_c} $, and $ {\bf{\bar h}}_{c,i}^H = \sum\nolimits_{j = 1}^K {{\bf{h}}_{c,j + \left( {i - 1} \right)K}^H}  $ is the $ i $th row of the group composite channel matrix $ {{{\bf{\bar H}}}_c} $, denoting the combined composite reflecting channel associated with the $ i $th IRS group. 
Based on this IRS elements grouping method, problem (\ref{P1}) is transformed into
\begin{subequations}\label{P1_1}
	\begin{align}
	\mathop {\max }\limits_{{\bf{Q}},{\bf{\bar v}}} & \ {\rm{ }}{\rm{ tr}}\left( {{\bf{P}}\left( {{\bf{\bar v}}} \right){\bf{Q}}} \right)\\
	{\rm{s}}{\rm{.t}}{\rm{. }} & \ \left| {{\bf{\bar v}}\left( n \right)} \right| = 1,\forall n \in {\cal J},\\
	& \ \text{(\ref{P1}c)}, \text{(\ref{P1}d)},
	\end{align}
\end{subequations}
where we define $ {\bf{P}}\left( {{\bf{\bar v}}} \right) = \eta {\bf{\bar H}}_c^H{\bf{\bar v}}{{{\bf{\bar v}}}^H}{{{\bf{\bar H}}}_c} $.

\subsection{Estimation of Group Composite Channel Matrix $ {{{\bf{\bar H}}}_c} $}
The following discussion is based on the assumption that for any $ {{\bf{\bar v}}} $, we can accurately estimate the scaled version of $ {\bf{P}}\left( {{\bf{\bar v}}} \right) $, which is denoted as $ {\bf{\tilde P}}\left( {{\bf{\bar v}}} \right) $. We will discuss it in detail in the next subsection. Since $ {\rm{rank}}\big( {{\bf{\tilde P}}\left( {{\bf{\bar v}}} \right)} \big) = 1 $, by performing the eigenvalue decomposition (EVD) of $ {\bf{\tilde P}}\left( {{\bf{\bar v}}} \right) $, we have $ {\bf{\tilde P}}\left( {{\bf{\bar v}}} \right) = {\bf{\tilde p}}\left( {{\bf{\bar v}}} \right){{{\bf{\tilde p}}}^H}\left( {{\bf{\bar v}}} \right) $, where $ \left\| {{\bf{\tilde p}}\left( {{\bf{\bar v}}} \right)} \right\| = \sqrt {{\rm{tr}}\big( {{\bf{\tilde P}}\left( {{\bf{\bar v}}} \right)} \big)} $. Accordingly, the following equation
\begin{align}\label{Effective_channel}
{\bf{\bar H}}_c^H{\bf{\bar v}} = \gamma {\bf{\tilde p}}\left( {{\bf{\bar v}}} \right)
\end{align}
holds, where $ \gamma $ is a complex scaling factor. Based on this, by generating $ C $ different IRS group reflection patterns, we have
\begin{align}
{\bf{\bar H}}_c^H{\bf{\bar V}} = {\bf{\Lambda }}{\bf{R}},
\end{align}
where $ {\bf{\bar V}} = \left[ {{{{\bf{\bar v}}}_1},{{{\bf{\bar v}}}_2}, \cdots ,{{{\bf{\bar v}}}_C}} \right] \in {\mathbb{C}^{J \times C}} $, $ {\bf{\Lambda }} = \left[ {{\bf{\tilde p}}\left( {{{{\bf{\bar v}}}_1}} \right),{\bf{\tilde p}}\left( {{{{\bf{\bar v}}}_2}} \right), \cdots ,{\bf{\tilde p}}\left( {{{{\bf{\bar v}}}_C}} \right)} \right] \in {\mathbb{C}^{{M_t} \times C}} $, and $ {\bf{R}} = {\rm{diag}}\left\{ {{\gamma _1},{\gamma _2}, \cdots ,{\gamma _C}} \right\} $. By applying the least squares (LS) estimator \cite{1597555_Biguesh}, the estimated $ {\bf{\bar H}}_c^{H,{\rm{LS}}} $ is given by
\begin{align}\label{LS_estimator}
{\bf{\bar H}}_c^{H,{\rm{LS}}} = {\bf{\Lambda }}{\bf{R}}{{{\bf{\bar V}}}^\dag },
\end{align}
where $ {{{\bf{\bar V}}}^\dag } = {{{\bf{\bar V}}}^H}{\left( {{\bf{\bar V}}{{{\bf{\bar V}}}^H}} \right)^{ - 1}} $ is the pseudoinverse of $ {{\bf{\bar V}}} $. It can be readily seen that $ C $ should satisfy $ C \ge J $. Therefore, to estimate the group composite channel $ {{{\bf{\bar H}}}_c} $, the key is to obtain the accurate estimate of the scaled version of $ \left\{ {{\bf{P}}\left( {{{{\bf{\bar v}}}_i}} \right)} \right\}_{i = 1}^C $, i.e., $ \big\{ {{\bf{\tilde P}}\left( {{{{\bf{\bar v}}}_i}} \right)} \big\}_{i = 1}^C $, and determine the values of $ \left\{ {{\gamma _i}} \right\}_{i = 1}^C $. In the next two subsections, we present the scaled channel learning approach and the scaling factor determination method, respectively.

\subsection{Scaled Channel Learning Using ACCPM}
In this subsection, with the given $ {{\bf{\bar v}}} $, we estimate the scaled version of $ {{\bf{P}}\left( {{\bf{\bar v}}} \right)} $, i.e., $ {\bf{\tilde P}}\left( {{\bf{\bar v}}} \right) $, based on the ER's one-bit feedback by applying the ACCPM \cite{6884811_Xu,7117443_Gopalakrishnan}. ACCPM is an efficient localization method to solve general convex or quasi-convex optimization problems, with the goal of finding a feasible point in a convex target set. Specifically, at each iteration, the algorithm computes an analytic center of the current working set defined by the cutting-planes generated in previous iterations. If the calculated analytic center is a solution, then the algorithm will terminate; otherwise a new cutting plane returned by the oracle will be added into the current working set, thus forming a new working set for the next iteration. As the number of iterations increases, the working set gradually shrinks, and the algorithm will eventually find a solution to the problem.

Recall that we only need to obtain an estimate of the scaled $ {\bf{P}}\left( {{\bf{\bar v}}} \right) $. For simplicity, we define the target set as
\begin{align}
\chi  \buildrel \Delta \over = \big\{ {{\bf{\tilde P}}\left( {{\bf{\bar v}}} \right)|{\bf{0}}  \preceq  {\bf{\tilde P}}\left( {{\bf{\bar v}}} \right) \preceq {\bf{I}},{\bf{\tilde P}}\left( {{\bf{\bar v}}} \right) = \beta {\bf{P}}\left( {{\bf{\bar v}}} \right),\forall \beta  > 0} \big\},
\end{align}
which contains all scaled matrices of $ {{\bf{P}}\left( {{\bf{\bar v}}} \right)} $ that satisfy $ {\bf{0}} \preceq \beta {\bf{P}}\left( {{\bf{\bar v}}} \right) \preceq {\bf{I}} $. Accordingly, the initial convex working set can be defined as $ {{\cal P}_0} \buildrel \Delta \over = \big\{ {{\bf{\tilde P}}\left( {{\bf{\bar v}}} \right)|{\bf{0}} \preceq {\bf{\tilde P}}\left( {{\bf{\bar v}}} \right) \preceq {\bf{I}}} \big\} $. Obviously, we have $ \chi  \subseteq {{\cal P}_0} $. In the following, we find a point in $ \chi  $ from $ {\cal P}{_0} $ by applying the ACCPM.

\begin{figure}[htbp]
	\centerline{\includegraphics[width=0.48\textwidth]{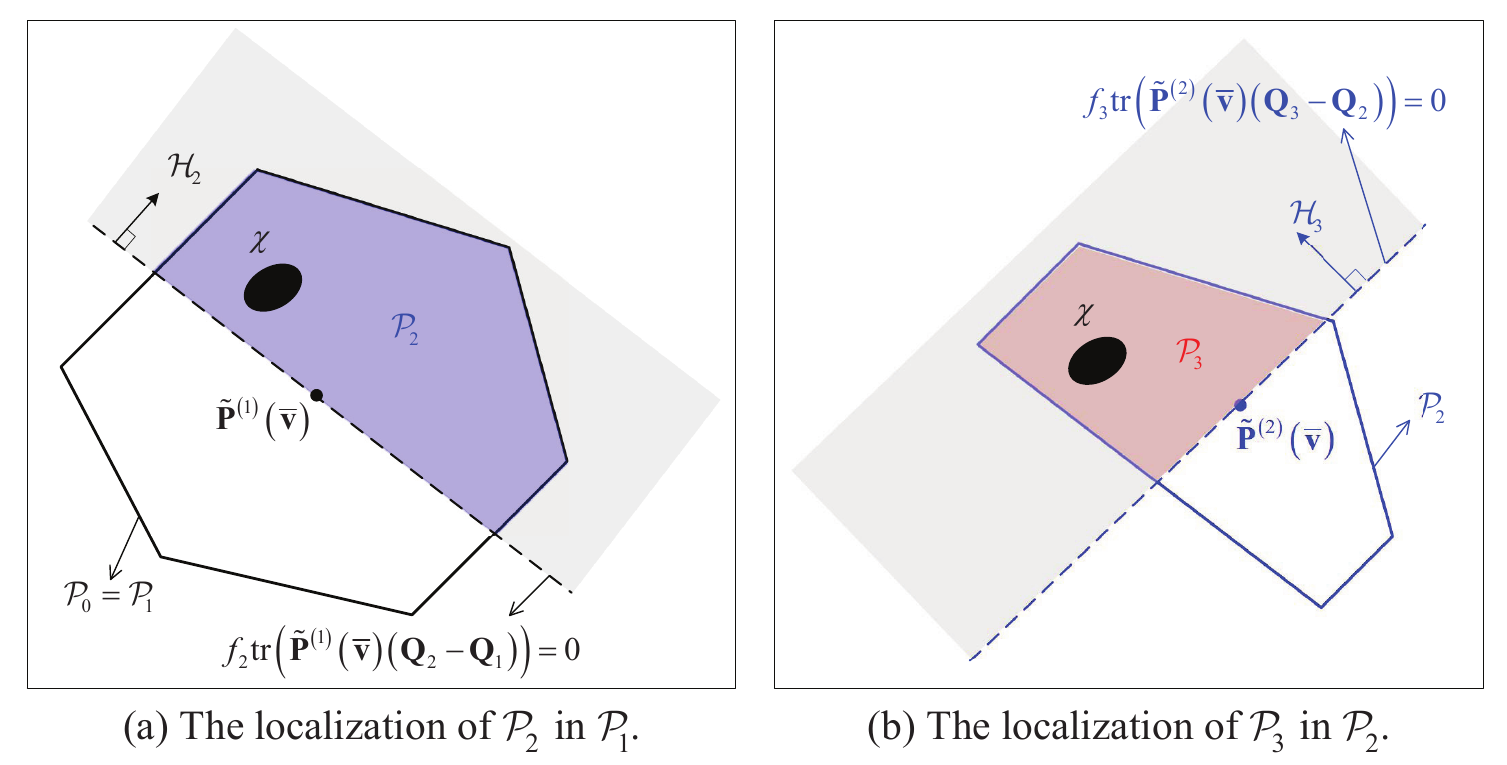}}
	\caption{The procedure of ACCPM: the working set gradually shrinks to the target set.}
	\label{fig_ACCPM}
\end{figure}

Denote the transmit covariance matrix at the ET in interval $ n, \forall n \in \mathbb{N} $, as $ {{\bf{Q}}_n} $. Then, the energy harvested at the ER over the $ n $th interval is given by
\begin{align}
{E_{h,n}} = {T_s}{\rm{tr}}\left( {{\bf{P}}\left( {{\bf{\bar v}}} \right){{\bf{Q}}_n}} \right),
\end{align}
where $ T_s $ is the duration of each interval. At the end of the interval $ n $, the ER measures the amount of its harvested energy $ E_{h,n} $, and feeds back one-bit information, denoted by $ {{\tilde f}_n} \in \left\{ {0,1} \right\} $, to indicate whether it is larger ($ {{\tilde f}_n} $ is set to `0') or smaller ($ {{\tilde f}_n} $ is set to `1') than that over the $ \left( n - 1 \right) $th interval. To facilitate our analysis, we set $ {f_n} = 2{{\tilde f}_n} - 1 $ such that $ {f_n} \in \left\{ { - 1,1} \right\} $. More concretely, we have $ {f_n} = -1 $ if $ {E_{h,n}} \ge {E_{h,n - 1}} $; or $ {f_n} = 1 $ otherwise. For the sake of completeness, we define $ {{\bf{Q}}_0} = {\bf{0}} $, such that $ {E_{h,0}} = 0 $. In this paper, we assume that $ \left\{ {{E_{h,n}}} \right\} $ are all perfectly measured at the ER. Therefore, each time the ET receives the one-bit feedback $ {f_n} $ from the ER, it can obtain the following inequality:
\begin{align}
{f_n}{\rm{tr}}\left( {{\bf{P}}\left( {{\bf{\bar v}}} \right)\left( {{{\bf{Q}}_n} - {{\bf{Q}}_{n - 1}}} \right)} \right) \le 0,\forall n \in \mathbb{N},
\end{align}
which means that $ {{\bf{\tilde P}}\left( {{\bf{\bar v}}} \right)} $ lies in the following half space:
\begin{align}
{{\cal H}_n} \buildrel \Delta \over = \left\{ {{\bf{S}}|{f_n}{\rm{tr}}\left( {{\bf{S}}\left( {{{\bf{Q}}_n} - {{\bf{Q}}_{n - 1}}} \right)} \right) \le 0} \right\}.
\end{align}
Accordingly, compared to the working set at interval $ n - 1 $, i.e., $ {\cal P}{_{n - 1}} $, the working set at interval $ n \left( n \ge 2 \right) $ is reduced to
\begin{align}
{{\cal P}_n} &= {{\cal P}_{n - 1}} \cap {{\cal H}_n} \notag\\
&= \big\{ {{\bf{\tilde P}}\left( {{\bf{\bar v}}} \right)|{\bf{0}} \preceq {\bf{\tilde P}}\left( {{\bf{\bar v}}} \right) \preceq {\bf{I}},} \big. \notag\\
&\qquad \quad \big. {{f_i}{\rm{tr}}\big( {{\bf{\tilde P}}\left( {{\bf{\bar v}}} \right)\left( {{{\bf{Q}}_i} - {{\bf{Q}}_{i - 1}}} \right)} \big) \le 0,2 \le i \le n} \big\}.
\end{align}
In particular, for $ n = 1 $, we set $ {{\cal P}_1} = {{\cal P}_0} $. After ER's $ N_f $ feedbacks, we have
\begin{align}
{{\cal P}_0} = {{\cal P}_1} \supseteq {{\cal P}_2} \supseteq  \cdots  \supseteq {{\cal P}_{{N_f}}} \supseteq \chi,
\end{align}
i.e., as the number of ER's feedbacks increases, the working set $ {{\cal P}_n} $ gradually shrinks to $ \chi  $. In Fig. \ref{fig_ACCPM}, we illustrate the procedures of locating $ {\cal P}{_{2}} $ in $ {\cal P}{_{1}} $ and locating $ {\cal P}{_{3}} $ in $ {\cal P}{_{2}} $.


Based on the principle of ACCPM, we need to query the oracle at the analytic center of the current working set $ {{\cal P}_n} $ for localizing the next working set $ {{\cal P}_{n + 1}} $. The analytic center of $ {{\cal P}_n} $ is explicitly given by \cite{sun2002analytic}
\begin{align}\label{Analytic_center}
{{{\bf{\tilde P}}}^{\left( n \right)}}\left( {{\bf{\bar v}}} \right) &= \arg {\rm{ }}\mathop {{\rm{min}}}\limits_{{\bf{0}} \preceq {\bf{\tilde P}} \preceq {\bf{I}}} {\rm{ }} - 2\log {\rm{ det}}\big( {{\bf{\tilde P}}} \big) - 2\log {\rm{ det}}\big( {{\bf{I}} - {\bf{\tilde P}}} \big) \notag\\
&- \sum\limits_{i = 2}^n {\log \big( { - {f_i}{\rm{tr}}\big( {{\bf{\tilde P}}\left( {{{\bf{Q}}_i} - {{\bf{Q}}_{i - 1}}} \right)} \big)} \big)} ,n \ge 0.
\end{align}
Although the problem in (\ref{Analytic_center}) is convex, its form does not support its direct solution with standard convex optimization solvers. For this reason, we rewrite the last term of (\ref{Analytic_center}) in a more compact form as follows.
Define $ {{{\bf{\bar Q}}}_i} =  - {f_i}\left( {{{\bf{Q}}_i} - {{\bf{Q}}_{i - 1}}} \right) $ and $ {\bf{\tilde Q}} = \left[ {{{{\bf{\bar Q}}}_2}, \cdots ,{{{\bf{\bar Q}}}_n}} \right] $, we have
\begin{align}\label{Analytic_center_1}
&\sum\limits_{i = 2}^n {\log \big( { - {f_i}{\rm{tr}}\big( {{\bf{\tilde P}}\left( {{{\bf{Q}}_i} - {{\bf{Q}}_{i - 1}}} \right)} \big)} \big)} \notag \\
&= {\bf{1}}_{n - 1}^T\log \big( {\left( {{{\bf{I}}_{n - 1}} \otimes {\bf{1}}_{{M_t}}^T} \right){\rm{diag}}\big( {\left( {{{\bf{1}}_{n - 1}} \otimes {{\bf{I}}_{{M_t}}}} \right){\bf{\tilde P\tilde Q}}} \big)} \big).
\end{align}
Combined with (\ref{Analytic_center_1}),
problem (\ref{Analytic_center}) can be efficiently solved by using standard convex optimization solvers such as CVX.

To efficiently implement the localization of $ \chi  $, with the given analytic center $ {{{\bf{\tilde P}}}^{\left( {n - 1} \right)}}\left( {{\bf{\bar v}}} \right) $ and the ET's transmit covariance matrix $ {{{\bf{Q}}_{n - 1}}} $ at interval $  n - 1  $, we need to guarantee that the resulting cutting plane is at least neutral, i.e., the cutting plane passes through the current analytic center. This guarantees a minimum worst-case total iteration number. Accordingly, the ET's transmit covariance matrix at interval $ n $, i.e., $ {{\bf{Q}}_n} $, is constructed to satisfy
\begin{align}
{\rm{tr}}\big( {{{{\bf{\tilde P}}}^{\left( n - 1 \right)}}\left( {{\bf{\bar v}}} \right)\left( {{{\bf{Q}}_n} - {{\bf{Q}}_{n - 1}}} \right)} \big) = 0,\forall n \ge 2.
\end{align}
By this way, with the initial ET's transmit covariance matrix for $ n = 1 $ set to $ {{\bf{Q}}_1} = \frac{{{P_b}}}{{{M_t}}}{{\bf{I}}_{{M_t}}} $, $ {\left\{ {{{\bf{Q}}_n}} \right\}_{n \ge 2}} $ can be obtained sequentially. Please refer to \cite{6884811_Xu} for details.

\subsection{Determination of Scale Factor $ \left\{ {{\gamma _i}} \right\} $}
We set the number of different IRS group reflection patterns to $ C = J $,
and design the IRS group reflection coefficient matrix $ {{\bf{\bar V}}} $ as the Hadamard matrix of order $ J $. Thus $ {{{\bf{\bar v}}}_j},\forall j \in {\cal J} $, is the $ j $th column of the $ J $-order Hadamard matrix. Since a Hadamard matrix of order $ 2 $ or $ 4 $$ k $ exists for every positive integer $ k $, we assume that $ J $ is a multiple of $ 4 $ without loss of generality. Thanks to the IRS elements grouping method, the number of groups can be appropriately adjusted to make it feasible. From (\ref{LS_estimator}), we have
\begin{align}\label{Estimation_Hc}
{\bf{\bar H}}_c^{H,{\rm{LS}}} &= {\bf{\Lambda R}}{{{\bf{\bar V}}}^\dag } \notag\\
&= {\bf{\Lambda R}}{{{\bf{\bar V}}}^T}{\left( {{\bf{\bar V}}{{{\bf{\bar V}}}^T}} \right)^{ - 1}} \notag\\
&\mathop  = \limits^{\left( a \right)} \frac{1}{J}{\bf{\Lambda R}}{{{\bf{\bar V}}}^T} \notag\\
&= \frac{1}{J}\sum\limits_{i = 1}^J {{\gamma _i}{\bf{\tilde p}}\left( {{{{\bf{\bar v}}}_i}} \right){\bf{\bar v}}_i^T},
\end{align}
where $ \left( a \right)  $ holds due to the property of Hadamard matrix. To perform joint active and passive beamforming design for the studied system, we only need to learn the scaled version of the cascaded channel $ {\bf{\bar H}}_c $. Therefore, we just need to determine the ratio between any two from $ \left\{ {{\gamma _i}} \right\}_{i = 1}^J $, including the amplitude ratio $ \left| {{\gamma _i}} \right|/\left| {{\gamma _j}} \right| $ and phase difference $ \angle {\gamma _i} - \angle {\gamma _j},\forall i \ne j $, and need not to know the true values of them. In the following, we convert the joint two-dimensional search for amplitude ratio and phase difference value into two separate one-dimensional searches. The details are given as follows. 

\subsubsection{Amplitude Ratio Determination} 
From (\ref{Effective_channel}), with the given IRS group reflection coefficients $ {{\bf{\bar v}}} $, the effective ET-ER channel can be expressed as
\begin{align}
{\bf{\tilde h}}_r^H = {{{\bf{\bar v}}}^H}{{{\bf{\bar H}}}_c} = {\gamma ^ * }{{{\bf{\tilde p}}}^H}\left( {{\bf{\bar v}}} \right).
\end{align}
By adopting the maximum ratio transmission (MRT) precoding scheme with the transmit power set to $ P_t $ at the ET, i.e., $ {\bf{x}} = \sqrt {{P_t}} \frac{{ {\bf{\tilde p}}\left( {{\bf{\bar v}}} \right)}}{{\left\| {{\bf{\tilde p}}\left( {{\bf{\bar v}}} \right)} \right\|}} $, the energy harvested at the ER over an interval is given by
\begin{align}\label{Eh_AmplitudeRatio}
{E_h}\left( {{P_t},{\bf{\bar v}}} \right) = \eta {T_s}{P_t}{\left| \gamma  \right|^2}{\left\| {{\bf{\tilde p}}\left( {{\bf{\bar v}}} \right)} \right\|^2}.
\end{align}
For any $ i \ne j $, we can always find $ {\bar P_{t,i}} $ and $ {\bar P_{t,j}} $ such that $ {E_{h,i}}\left( {{{{\bar P}_{t,i}}},{{{\bf{\bar v}}}_{i}}} \right) = {E_{h,j}}\left( {{{{\bar P}_{t,j}}},{{{\bf{\bar v}}}_{j}}} \right) $ based on the ER's one-bit feedback by applying the one-dimensional search in the power dimension. Then, the amplitude ratio can be calculated as
\begin{align}\label{Amplitude_ratio}
\frac{{\left| {{\gamma _i}} \right|}}{{\left| {{\gamma _j}} \right|}} = \sqrt {\frac{{{{{\bar P}_{t,j}}}}}{{{{{\bar P}_{t,i}}}}}} \frac{{\left\| {{\bf{\tilde p}}\left( {{{{\bf{\bar v}}}_{j}}} \right)} \right\|}}{{\left\| {{\bf{\tilde p}}\left( {{{{\bf{\bar v}}}_{i}}} \right)} \right\|}}, \forall i \ne j.
\end{align} 
The procedures for finding $ {\bar P_{t,i}} $ and $ {\bar P_{t,j}} $ to satisfy $ {E_{h,i}}\left( {{{{\bar P}_{t,i}}},{{{\bf{\bar v}}}_{i}}} \right) = {E_{h,j}}\left( {{{{\bar P}_{t,j}}},{{{\bf{\bar v}}}_{j}}} \right) $ are given as follows.
Denote by the system configuration $ \left\{ {{P_t},{\bf{\bar v}}} \right\} $ as that the precoding vector $ {\bf{x}} = \sqrt {{P_t}} \frac{{ {\bf{\tilde p}}\left( {{\bf{\bar v}}} \right)}}{{\left\| {{\bf{\tilde p}}\left( {{\bf{\bar v}}} \right)} \right\|}} $ is adopted at the ET, and the group reflection coefficients $ {{\bf{\bar v}}} $ is set at the IRS.
In the first two timeslots, the system is configured with $ \left\{ {{P_b},{{{\bf{\bar v}}}_{i}}} \right\} $ and $ \left\{ {{P_b},{{{\bf{\bar v}}}_{j}}} \right\} $, respectively. Without loss of generality, we assume that $ {E_h}\left( { {{P_b},{{{\bf{\bar v}}}_{i}}} } \right) > {E_h}\left( { {{P_b},{{{\bf{\bar v}}}_{j}}} } \right) $. Then, in the subsequent timeslots, we fix $ {P_{t,j}} = {P_b} $, and use the bisection method to search for $ {{P_{t,i}}} $. By alternately configuring $ \left\{ {{P_{t,i}},{{{\bf{\bar v}}}_{i}}} \right\} $ and $ \left\{ {{P_b},{{{\bf{\bar v}}}_{j}}} \right\} $, the $ {{{\bar P}_{t,i}}} $ can be eventually obtained such that $ {E_{h,i}}\big( {{{\bar P}_{t,i}},{{{\bf{\bar v}}}_{i}}} \big) = {E_{h,j}}\left( {{P_b},{{{\bf{\bar v}}}_{j}}} \right) $.

\subsubsection{Phase Difference Determination} Prior to proceeding it, for any $ i \ne j $, we multiply the vector $ {\bf{\tilde p}}\left( {{{{\bf{\bar v}}}_{j}}} \right) $ by a phase factor $ e^{j{\varpi _j}} $, i.e., $ {\bf{\tilde p}}\left( {{{{\bf{\bar v}}}_{j}}} \right) \leftarrow {e^{j{\varpi _j}}}{\bf{\tilde p}}\left( {{{{\bf{\bar v}}}_{j}}} \right) $, such that $ \Im  \left\{ {{{{\bf{\tilde p}}}^H}\left( {{{{\bf{\bar v}}}_{i}}} \right){\bf{\tilde p}}\left( {{{{\bf{\bar v}}}_{j}}} \right)} \right\} = 0 $.
Then, by setting the IRS group reflection coefficients as\footnote{The value of each element in $ {\bf{\bar v}}_{i,j}^c $ takes from $ \left\{ { - 1,0,1} \right\} $, where $ 0 $ means that the corresponding IRS elements are switched off \cite{9130088_Wang}.}
\begin{align}
{\bf{\bar v}}_{i,j}^c = \frac{1}{2}\left( {{{{\bf{\bar v}}}_{i}} + {{{\bf{\bar v}}}_{j}}} \right),
\end{align}
we have
\begin{align}
&{\bf{\tilde h}}_r^H\left( {{\bf{\bar v}}_{i,j}^c} \right) = {\bf{\bar v}}_{i,j}^{c,T}{\bf{\bar H}}_c^{{\rm{LS}}} \notag \\
&= \frac{1}{{2J}}\left( {{\bf{\bar v}}_{i}^T + {\bf{\bar v}}_{j}^T} \right)\sum\limits_{m = 1}^J {\gamma _m^ * {{{\bf{\bar v}}}_{m}}{{{\bf{\tilde p}}}^H}\left( {{{{\bf{\bar v}}}_{m}}} \right)} \notag\\
&= \frac{1}{2}\left( {\gamma _i^ * {{{\bf{\tilde p}}}^H}\left( {{{{\bf{\bar v}}}_{i}}} \right) + \gamma _j^ * {{{\bf{\tilde p}}}^H}\left( {{{{\bf{\bar v}}}_{j}}} \right)} \right) \notag\\
& = \frac{1}{2}\left| {{\gamma _j}} \right|{e^{ - j\angle {\gamma _j}}}\big( {\frac{{\left| {{\gamma _i}} \right|}}{{\left| {{\gamma _j}} \right|}}{e^{ - j\left( {\angle {\gamma _i} - \angle {\gamma _j}} \right)}}{{{\bf{\tilde p}}}^H}\left( {{{{\bf{\bar v}}}_{i}}} \right) + {{{\bf{\tilde p}}}^H}\left( {{{{\bf{\bar v}}}_{j}}} \right)} \big).
\end{align}
The following proposition shows that the phase difference $ \angle {\gamma _i} - \angle {\gamma _j} $ can be determined by applying the one-dimensional search method.

\newtheorem{theorem}{Proposition}
\begin{theorem}\label{Proposition_Phase_difference}
	Let the transmit beamforming vector be
	\begin{align}
	{{\bf{x}}_{i,j}}\left( \vartheta  \right) = \sqrt {{P_b}} \frac{{\left| {{\gamma _i}} \right|/\left| {{\gamma _j}} \right|{e^{j\vartheta }}{\bf{\tilde p}}\left( {{{{\bf{\bar v}}}_{i}}} \right) + {\bf{\tilde p}}\left( {{{{\bf{\bar v}}}_{j}}} \right)}}{{\left\| {\left| {{\gamma _i}} \right|/\left| {{\gamma _j}} \right|{e^{j\vartheta }}{\bf{\tilde p}}\left( {{{{\bf{\bar v}}}_{i}}} \right) + {\bf{\tilde p}}\left( {{{{\bf{\bar v}}}_{j}}} \right)} \right\|}}.
	\end{align}
	Then, the value of $ \angle {\gamma _i} - \angle {\gamma _j} $ can be found by
	\begin{align}
	\angle {\gamma _i} - \angle {\gamma _j} = \arg \mathop {{\rm{ext}}}\limits_\vartheta \  {E_h}\left( \vartheta  \right) \buildrel \Delta \over = \eta {T_s}{\big| {{\bf{\tilde h}}_r^H\left( {{\bf{\bar v}}_{i,j}^c} \right){{\bf{x}}_{i,j}}\left( \vartheta  \right)} \big|^2},
	\end{align}
	where we use $ \arg {\rm{ ext}} $ to denote an argument that yields the unique local maximum point within the interval of $ \left[ {0,2\pi } \right) $.
\end{theorem}

\begin{IEEEproof}
	Please see Appendix \ref{appendix_A}.
\end{IEEEproof}

Combining the searched amplitude ratio $ \left| {{\gamma _i}} \right|/\left| {{\gamma _j}} \right| $ and phase difference $ \angle {\gamma _i} - \angle {\gamma _j},\forall i \ne j $, the estimate of the scaled version of $ {\bf{\bar H}}_c $ can be obtained. The proposed one-bit-feedback-based cascaded channel estimation method for the studied IRS-aided WET system is summarized in Algorithm \ref{alg:AO}.

\begin{algorithm}
	\caption{One-bit-feedback-based channel estimation method for the IRS-aided WET system.}
	\label{alg:AO}
	\begin{algorithmic}[1]
		{\begin{small}
				\STATE Initialize the IRS elements grouping, and denote the group number as $ J $.
				\FOR{$ i = 1,2, \cdots ,J $}
				\STATE Set the IRS group reflection coefficients $ {{\bf{\bar v}}} $ as the $ i $th column of the $ J $-order Hadamard matrix, and denote it as $ {{{{\bf{\bar v}}}_i}} $.
				\STATE Estimate the scaled version of the effective ET-ER channel with the given $ {{{{\bf{\bar v}}}_i}} $, i.e., $ {\bf{\bar v}}_i^H{{{\bf{\bar H}}}_c} $, by applying the ACCPM, and denote it as $ {{{\bf{\tilde p}}}^H}\left( {{{{\bf{\bar v}}}_i}} \right) $.
				\ENDFOR
				\FOR{$ j = 2,3, \cdots ,J $}
				\STATE Obtain $ \left| {{\gamma _j}} \right|/\left| {{\gamma _1}} \right| $ according to (\ref{Amplitude_ratio}) by applying the one-dimensional search method.
				\STATE Obtain $ \angle {\gamma _j} - \angle {\gamma _1} $ according to Proposition \ref{Proposition_Phase_difference} by applying the one-dimensional search method.
				\ENDFOR
				\STATE Let $ {\gamma _1} = 1 $, and calculate $ \left\{ {{\gamma _j}} \right\}_{j = 2}^J $. The scaled version of the cascaded channel matrix $ {\bf{\bar H}}_c $ is given by $ {\bf{\bar H}}_c^{H,{\rm{LS}}} = \sum\nolimits_{i = 1}^J {{\gamma _i}{\bf{\tilde p}}\left( {{{{\bf{\bar v}}}_i}} \right){\bf{\bar v}}_i^T}  $.
		\end{small}}
	\end{algorithmic}
\end{algorithm}

\subsection{Downlink Transmission Design}
Based on the estimated cascaded channel $ {\bf{\bar H}}_c^{{\rm{LS}}} $, we can jointly optimize the ET's transmit covariance matrix $ {\bf{Q}} $ and the IRS group reflection coefficients $ {{\bf{\bar v}}} $ by solving the following problem to maximize the amount of the energy harvested at the ER, i.e.,
\begin{subequations}\label{P2}
\begin{align}
\mathop {\max }\limits_{{\bf{Q}},{\bf{\bar v}}} & \ \eta {\rm{tr}}\big( {{{{\bf{\bar v}}}^H}{\bf{\bar H}}_c^{{\rm{LS}}}{\bf{Q\bar H}}_c^{{\rm{LS}},H}{\bf{\bar v}}} \big)\\
{\rm{s}}{\rm{.t}}{\rm{. }}
& \ \left( {\text{\ref{P1_1}}{\text{b}}} \right), \left( {\text{\ref{P1}}{\text{c}}} \right), \left( {\text{\ref{P1}}{\text{d}}} \right).
\end{align}
\end{subequations}
The following proposition helps to derive a closed form optimal $ {\bf{Q}} $ for problem (\ref{P2}).

\begin{theorem}\label{Proposi_optimal_Q}
	For the following optimization problem
	\begin{subequations}
		\begin{align}
		\mathop {\max }\limits_{\bf{Q}} & \ {\rm{ tr}}\left( {{\bf{QU}}} \right)\\
		{\rm{s}}{\rm{.t}}{\rm{.}} & \ {\rm{tr}}\left( {\bf{Q}} \right) \le {P_b},{\bf{Q}} \succeq {\bf{0}},
		\end{align}
	\end{subequations}	
	where the matrix $ \bf{U} $ has the same dimensions as $ \bf{Q} $. The optimal $ {\bf{Q}} $ is given by $ {{\bf{Q}}^{{\rm{opt}}}} = {P_b}{{\bm{\upsilon }}_U}{\bm{\upsilon }}_U^H $, where $ {{\bm{\upsilon }}_U} $ is the eigenvector of $ \bf{U} $ corresponding to its dominant eigenvalue.
\end{theorem}

\begin{IEEEproof}
	Please refer to Proposition 2.1 in \cite{6489506_Zhang}.
\end{IEEEproof}

According to Proposition \ref{Proposi_optimal_Q}, the optimal $ {\bf{Q}} $ for problem (\ref{P2}) is given by $ {{\bf{Q}}^{{\rm{opt}}}} = {P_b}\frac{{{\bf{\bar H}}_c^{{\text{LS}},H}{\bf{\bar v}}{{{\bf{\bar v}}}^H}{\bf{\bar H}}_c^{{\text{LS}}}}}{{{{\left\| {{\bf{\bar H}}_c^{{\text{LS}},H}{\bf{\bar v}}} \right\|}^2}}} $.
Substituting the expression of $ {{\bf{Q}}^{{\rm{opt}}}} $ into problem (\ref{P2}) yields the following optimization problem with respect to $ {\bf{\bar v}} $:
\begin{subequations}
	\begin{align}
	\mathop {\max }\limits_{{\bf{\bar v}}} & \ \eta {P_b}{{{\bf{\bar v}}}^H}{\bf{\bar H}}_c^{{\rm{LS}}}{\bf{\bar H}}_c^{{\rm{LS}},H}{\bf{\bar v}}\\
	{\rm{s}}{\rm{.t}}{\rm{. }}& \ \left( \text{{\ref{P1_1}b}} \right),
	\end{align}
\end{subequations} 
which can be efficiently solved by applying a variety of optimization methods such as element-wise block coordinate descent (BCD), relaxation and projection, semidefinite relaxation (SDR), etc. \cite{9847080_Pan}.

\subsection{Feedback Number Complexity Analysis}
The number of one-bit feedbacks from the ER to the ET determines the accuracy of the estimated $ {{\bf{\bar H}}_c} $. Here, we briefly analyze the number of feedbacks required for the given ET's transmit antenna number $ M_t $ and IRS group number $ J $. According to Proposition 3.2 in \cite{6884811_Xu}, the number of feedbacks required for the ACCPM is at most $ {\cal O}\big( {\frac{{M_t^3}}{{{\epsilon^2}}}} \big) $, where $ \epsilon $ is the accuracy of $ {\bf{\tilde P}}\left( {{\bf{\bar v}}} \right) $. From steps 2-5 of Algorithm 1, the number of the ACCPM invoking is $ J $. Moreover, the number of feedbacks for steps 6-9 is also proportional to $ J $. Therefore, the feedback number complexity of the proposed one-bit-feedback-based channel estimation method is $ {\cal O}\left( {JM_t^3} \right) $.

\section{Distributed-Beamforming-Based Method}
In this section, we propose a joint active ET and passive IRS beamforming design method based on the ER's one-bit feedback, which does not require the estimation of the cascaded ET-IRS-ER channel. Specifically, we first leverage the distributed transmit beamforming method in \cite{5361473_Mudumbai} based on one-bit feedback to optimize the IRS (group) reflection coefficients. Then, with the optimized IRS (group) reflection coefficients, we only need to invoke the ACCPM once to obtain the corresponding optimal ET's transmit covariance matrix.

To speed up the convergence of the passive IRS beamforming, we still adopt the IRS elements grouping method in Section III.
According to Proposition \ref{Proposi_optimal_Q}, the optimal ET's transmit covariance matrix $ \bf{Q} $ for problem (\ref{P1_1}) is given by $ {{\bf{Q}}^{{\rm{opt}}}} = {P_b}\frac{{{\bf{\bar H}}_c^H{\bf{\bar v}}{{{\bf{\bar v}}}^H}{{{\bf{\bar H}}}_c}}}{{{{\left\| {{\bf{\bar H}}_c^H{\bf{\bar v}}} \right\|}^2}}} $. Substituting $ {{\bf{Q}}^{{\rm{opt}}}} $ into problem (\ref{P1_1}) yields the following problem
\begin{subequations}\label{P3}
	\begin{align}
	\mathop {\max }\limits_{{\bf{\bar v}}} \ & {\rm{ }}\eta {P_b}{{{\bf{\bar v}}}^H}{{{\bf{\bar H}}}_c}{\bf{\bar H}}_c^H{\bf{\bar v}}\\
	{\rm{s}}{\rm{.t}}{\rm{. }} \ & \left( \text{{\ref{P1_1}b}} \right).
	\end{align}
\end{subequations}
Denote the subproblem with respect to $ {{\bf{\bar v}}} $ resulted by substituting $ {\bf{Q}} = \frac{{{P_b}}}{{{M_t}}}{{\bf{I}}_{{M_t}}} $ into problem (\ref{P1_1}) as problem (7').
It can be readily seen that problems (7') and (\ref{P3}) share the same optimal $ {{\bf{\bar v}}} $. Based on this, to obtain the optimal $ {{\bf{\bar v}}} $ for problem (\ref{P1_1}), we can first optimize $ {{\bf{\bar v}}} $ with $ {\bf{Q}} $ set to $ \frac{{{P_b}}}{{{M_t}}}{{\bf{I}}_{{M_t}}} $, which leads to the following problem
\begin{subequations}\label{P4}
	\begin{align}
	\mathop {\max }\limits_{{\bf{\bar v}}} & \ {\rm{ }}{E_Q}\left( {{\bf{\bar v}}} \right) = {{{\bf{\bar v}}}^H}{\bf{\Phi \bar v}}\\
	{\rm{s}}{\rm{.t}}{\rm{. }} & \ \left( \text{{\ref{P1_1}b}} \right),
	\end{align}
\end{subequations}
where $ {\bf{\Phi }} = \eta \frac{{{P_b}}}{{{M_t}}}{{{\bf{\bar H}}}_c}{\bf{\bar H}}_c^H $ is a deterministic but unknown quantity. Suppose that we have the optimized $ {{\bf{\bar v}}} $ for problem (\ref{P4}), then the scaled $ {\bf{P}}\left( {{\bf{\bar v}}} \right) $ (recall that $ {\bf{P}}\left( {{\bf{\bar v}}} \right) = \eta {\bf{\bar H}}_c^H{\bf{\bar v}}{{\bf{\bar v}}^H}{{\bf{\bar H}}_c} $), i.e., $ {\bf{\tilde P}}\left( {{\bf{\bar v}}} \right) $, can be obtained by applying the ACCPM. Further, with the given scaled $ {\bf{P}}\left( {{\bf{\bar v}}} \right) $, the optimal $ \bf{Q} $ can be accordingly obtained by solving the subproblem with respect to $ \bf{Q} $ in problem (\ref{P1_1}), which is expressed as $ {{\bf{Q}}^{{\rm{opt}}}} = {P_b}{{\bm{\upsilon }}_P}{\bm{\upsilon }}_P^H $ according to Proposition \ref{Proposi_optimal_Q}, where $ \bm{\upsilon }_P $ is the eigenvector of $ {\bf{\tilde P}}\left( {{\bf{\bar v}}} \right) $ corresponding to its dominant eigenvalue. Therefore, in the following content we will mainly focus on the optimization of $ {{\bf{\bar v}}} $ in problem (\ref{P4}) without knowing the matrix $ {\bf{\Phi }} $.

We borrow the idea of distributed transmit beamforming using one-bit feedback in \cite{5361473_Mudumbai} to optimize the IRS group reflection coefficients $ {{\bf{\bar v}}} $. 
Distributed beamforming is a form of collaborative communication in which multiple information sources simultaneously send a common message and control the phase of their transmissions to make the signals add constructively at an intended destination \cite{5361473_Mudumbai,4785387_Mudumbai}.
Accordingly, each group of IRS elements can be treated as an individual information source.
The adaptation is also performed in time-slotted fashion, with each IRS group reflection coefficient adapted in a timeslot in response to the feedback from the ER. Specifically, at the beginning of timeslot $ n $, denote the best known $ i $th IRS group PS as $ {{\bar \theta }_i}\left[ n \right] $. Then, at timeslot $ n $, a random phase perturbation $ {\delta _i}\left[ n \right] $ is applied to each IRS group PS $ {{\bar \theta }_i}\left[ n \right] $ to probe for a potentially better IRS reflection coefficient. Thus, the ``probe" IRS group PSs in timeslot $ n $ is given by
\begin{align}
{{{\bm{\bar \theta }}}^{{\rm{probe}}}}\left[ n \right] = {\bm{\bar \theta }}\left[ n \right] + {\bm{\delta }}\left[ n \right],
\end{align}
where $ {{{\bm{\bar \theta }}}^{{\rm{probe}}}}\left[ n \right] = {\big[ {\bar \theta _1^{{\rm{probe}}}\left[ n \right],\bar \theta _2^{{\rm{probe}}}\left[ n \right], \cdots ,\bar \theta _J^{{\rm{probe}}}\left[ n \right]} \big]^T} $, $ {\bm{\bar \theta }}\left[ n \right] = {\left[ {{{\bar \theta }_1}\left[ n \right],{{\bar \theta }_2}\left[ n \right], \cdots ,{{\bar \theta }_J}\left[ n \right]} \right]^T} $, and $ {\bm{\delta }}\left[ n \right] = {\left[ {{\delta _1}\left[ n \right],{\delta _2}\left[ n \right], \cdots ,{\delta _J}\left[ n \right]} \right]^T} $. Accordingly, the energy harvested at the ER over the $ n $th timeslot is given by
\begin{align}
Y\left[ n \right] = {{{\bf{\bar v}}}^H}\left( {{{{\bm{\bar \theta }}}^{{\rm{probe}}}}\left[ n \right]} \right){\bf{\Phi \bar v}}\left( {{{{\bm{\bar \theta }}}^{{\rm{probe}}}}\left[ n \right]} \right).
\end{align}
The ER measures $ Y\left[ n \right] $, and then feeds back one-bit information to the ET indicating whether $ Y\left[ n \right] $ is larger or smaller than its recorded maximum amount of harvested energy so far, which is given by
\begin{align}
{Y_{{\rm{best}}}}\left[ n \right] \buildrel \Delta \over = \mathop {\max }\limits_{m < n} {\rm{ }} \ Y\left[ m \right].
\end{align}
If the feedback information from the ER indicates an increase in the harvested energy amount, then the IRS keeps its random phase perturbations. Otherwise it undoes the phase perturbation. Therefore, the best known IRS group PSs at the beginning of timeslot $ \left(n + 1 \right) $ are updated as
\begin{align}\label{Update_theta}
{{\bar \theta }_i}\left[ {n + 1} \right] = \left\{ {\begin{array}{*{20}{c}}
	{{{\bar \theta }_i}\left[ n \right] + {\delta _i}\left[ n \right],{\text{ if }}Y\left[ n \right] > {Y_{{\rm{best}}}}\left[ n \right],}\\
	{{{\bar \theta }_i}\left[ n \right],{\text{ otherwise,}}}
	\end{array}} \right.\forall i \in {\cal J}.
\end{align}
Moreover, the ER also updates the recorded maximum amount of harvested energy so far as
\begin{align}
{Y_{{\rm{best}}}}\left[ {n + 1} \right] = \max \left\{ {{Y_{{\rm{best}}}}\left[ n \right],Y\left[ n \right]} \right\}.
\end{align}
The above perturbation-based exploration procedure for better IRS group reflection coefficients is repeated over multiple timeslots. Since the update strategy for IRS group PSs, i.e., (\ref{Update_theta}), ensures that the amount of energy harvested at the ER is monotonically non-decreasing over timeslots, and the optimal value of problem (\ref{P4}) has an upper bound, then the convergence will eventually be achieved. Further, we provide an argument that for arbitrary initial IRS group PSs $ {\bm{\bar \theta }}\left[ 0 \right] $, $ {\left\{ {{Y_{{\rm{best}}}}\left[ n \right]} \right\}_{n \in \mathbb{N}}} $ converges almost surely to a local maximum of the function $ {E_Q}\big( {{\bf{\bar v}}\big( {{\bm{\bar \theta }}} \big)} \big) $ in (\ref{P4}a), which is denoted as $ {{Y_{{\rm{opt}}}}} $. Prior to proving it, we present the following proposition stating that as long as $ {\bm{\bar \theta }}\left[ n \right] $ is not at the local maximum of $ {E_Q}\big( {{\bf{\bar v}}\big( {{\bm{\bar \theta }}} \big)} \big) $, there is always a finite probability of obtaining a finite increase in $ {E_Q} $ in each subsequent timeslot, which will be used to establish the convergence.

\begin{theorem}\label{Proposition_IncreasePro}
	Suppose that the probability density function (PDF) of the random phase perturbation $ \delta_i $, i.e., $ g\left( {{\delta _i}} \right) $, is bounded away from zero over an interval $ \left( { - {\Delta _0},{\Delta _0}} \right) $, where $ {\Delta _0} > 0 $. Then, for any $ \epsilon > 0 $, there exist positive $ \epsilon_1 $ and $ \rho $ such that
	\begin{align}
	&{E_Q}\left( {{\bf{\bar v}}\left( {{\bm{\bar \theta }}} \right)} \right) \le {Y_{{\rm{opt}}}} - \epsilon \Rightarrow \notag\\
	&\Pr \left( {{Y_{{\rm{best}}}}\left[ {n + 1} \right] - {Y_{{\rm{best}}}}\left[ n \right] \ge {\epsilon_1}|{\bm{\bar \theta }}\left[ n \right] = {\bm{\bar \theta }}} \right) \ge \rho ,\forall {\bm{\bar \theta }}.
	\end{align}
\end{theorem}

\begin{IEEEproof}
	Please see Appendix \ref{appendix_B}.
\end{IEEEproof}

\begin{theorem}\label{Theorem}
	For the PDF $ g\left( {{\delta _i}} \right) $ considered in Proposition~\ref{Proposition_IncreasePro}, given an arbitrary initial $ {\bm{\bar \theta }}\left[ 0 \right] $, the distributed beamforming algorithm will finally converge to a local maximum of the function $ {E_Q}\big( {{\bf{\bar v}}\big( {{\bm{\bar \theta }}} \big)} \big) $ almost surely, i.e., $ {Y_{{\rm{best}}}}\left[ n \right] \to {Y_{{\rm{opt}}}} $ or equivalently $ \frac{{\partial {E_Q}\left( {{\bf{\bar v}}\left( {{\bm{\bar \theta }}} \right)} \right)}}{{\partial {{\bar \theta }_i}}} = 0,\forall i \in {\cal J} $, with probability 1.
\end{theorem}

\begin{IEEEproof}
We prove the theorem by contradiction. For some $ \epsilon > 0 $, according to Proposition \ref{Proposition_IncreasePro}, we can conclude that there exists some $ \mu > 0 $ such that
\begin{align}\label{Theorem_equ_1}
&{E_Q}\left( {{\bf{\bar v}}\left( {{\bm{\bar \theta }}} \right)} \right) \le {Y_{{\rm{opt}}}} - \epsilon \Rightarrow \notag\\
&\mathbb{E}\left\{ {{Y_{{\rm{best}}}}\left[ {n + 1} \right] - {Y_{{\rm{best}}}}\left[ n \right]|{\bm{\bar \theta }}\left[ n \right] = {\bm{\bar \theta }}} \right\} \ge \mu ,\forall {\bm{\bar \theta }}.
\end{align}	
In particular, we can take $ \mu = \rho\epsilon_1 $ to satisfy (\ref{Theorem_equ_1}). Suppose that there exists some $ \epsilon' > 0 $ such that when $ n $ goes to infinite, the following inequality
\begin{align}
{E_Q}\left( {{\bf{\bar v}}\left( {{\bm{\bar \theta }}}{\left[ n \right]} \right)} \right) \le {Y_{{\rm{opt}}}} - \epsilon'
\end{align}
still holds. Then, for some $ \mu'\left( n \right) > 0 $, we have 
\begin{align}
\mathbb{E}\left\{ {{Y_{{\rm{best}}}}\left[ {n + 1} \right] - {Y_{{\rm{best}}}}\left[ n \right]} \right\} \ge \mu '\left( n \right),n \to \infty .
\end{align}
Accordingly, with probability one, we have
\begin{align}\label{Theorem_equ_2}
&{Y_{{\rm{opt}}}} \ge \mathbb{E}\big\{ {\mathop {\lim }\limits_{n \to \infty } {\rm{ }}{Y_{{\rm{best}}}}\left[ n \right]} \big\} \notag\\
&= \mathbb{E}\left\{ {{Y_{{\rm{best}}}}\left[ 0 \right]} \right\} + \mathbb{E}\big\{ {\sum\limits_{n = 0}^\infty  {{Y_{{\rm{best}}}}\left[ {n + 1} \right] - {Y_{{\rm{best}}}}\left[ n \right]} } \big\} \notag\\
&= \mathbb{E}\left\{ {{Y_{{\rm{best}}}}\left[ 0 \right]} \right\} + \sum\limits_{n = 0}^\infty  \mathbb{E}{\left\{ {{Y_{{\rm{best}}}}\left[ {n + 1} \right] - {Y_{{\rm{best}}}}\left[ n \right]} \right\}} \notag \\
&\ge \mathbb{E}\left\{ {{Y_{{\rm{best}}}}\left[ 0 \right]} \right\} + \sum\limits_{n = 0}^\infty  {\mu '\left( n \right)} \to \infty.
\end{align}
Since any $ {Y_{{\rm{opt}}}} $ is a finite positive number, which contradicts $ {Y_{{\rm{opt}}}} \to \infty $ in (\ref{Theorem_equ_2}). Accordingly, we conclude that the hypothetical $ \epsilon' > 0  $ for $ n \to \infty $ does not exist. This completes the proof.
\end{IEEEproof}

We now analyze the number of one-bit feedbacks required for the proposed distributed-beamforming-based method. It can be readily seen that for the first stage of the IRS group PSs optimization, the number of feedbacks required is only related to the number of IRS elements groups $ J $. More accurately, the required feedback number is sublinear with respect to $ J $, i.e., no more than $ {\cal O}\left( J \right) $, from the simulation results. For the second stage of the effective ET-ER channel learning, since we only need to invoke the ACCPM once, the number of feedbacks required is approximately $ {\cal O}\left( {M_t^3} \right) $. Therefore, the total number of feedbacks required for the distributed-beamforming-based method is at most $ {\cal O}\left( {J + M_t^3} \right) $.

\begin{figure}[htbp]
	\centerline{\includegraphics[width=0.38\textwidth]{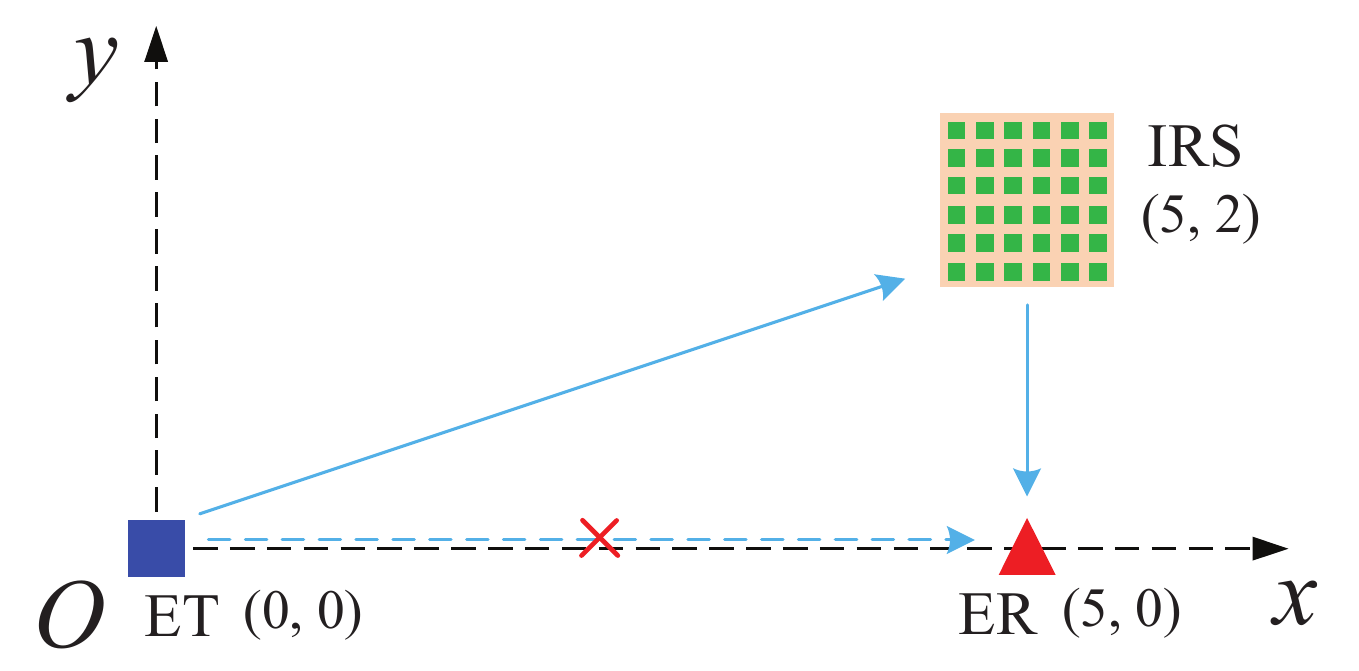}}
	\caption{Simulation setup.}
	\label{fig_system_setup}
\end{figure}

\section{Numerical Results}
In this section, numerical results are provided to demonstrate the effectiveness of our proposed joint active and passive beamforming designs exploiting one-bit feedback for our studied IRS-aided WET system. A two-dimensional (2D) coordinate setup is considered where the ET, IRS, and ER are located at $ \left( {{\text{0 m}},{\text{0 m}}} \right) $, $ \left( {{\text{5 m}},{\text{2 m}}} \right) $, and $ \left( {{\text{5 m}},{\text{0 m}}} \right) $, respectively. Both ET-IRS channel and IRS-ER channel are assumed to include large-scale fading and small-scale fading. The large-scale path loss is expressed as $ L\left( d \right) = {L_0}{\left( {d/{d_0}} \right)^{ - \alpha }} $, where $ d $ is the link distance, $ {L_0} =  - 30{\text{ dB}} $ represents the channel power gain at the reference distance $ {d_0} = 1{\text{ m}} $, and $ \alpha = 2.2 $ is the path loss exponent. To account for small-scale fading, we assume that both ET-IRS and IRS-ER channels follow Rayleigh fading for simplicity. For the distributed-beamforming-based method, the random perturbations are sampled independently and uniformly from $ \left( { - {{\bar \Delta }_0},{{\bar \Delta }_0}} \right)  $, i.e., $ g\left( {{\delta _i}} \right) = \frac{1}{{2{{\bar \Delta }_0}}} $. 
All the results are averaged over 10 independent channel realizations except for the convergence behaviour of the ACCPM.
Unless otherwise specified, we set $ {P_b} = 30{\text{ dBm}} $, $ \eta = 0.5 $, and $ N = 64 $.

\subsection{Channel-Estimation-Based Method}
\begin{figure}[htbp]
	\centerline{\includegraphics[width=0.4\textwidth]{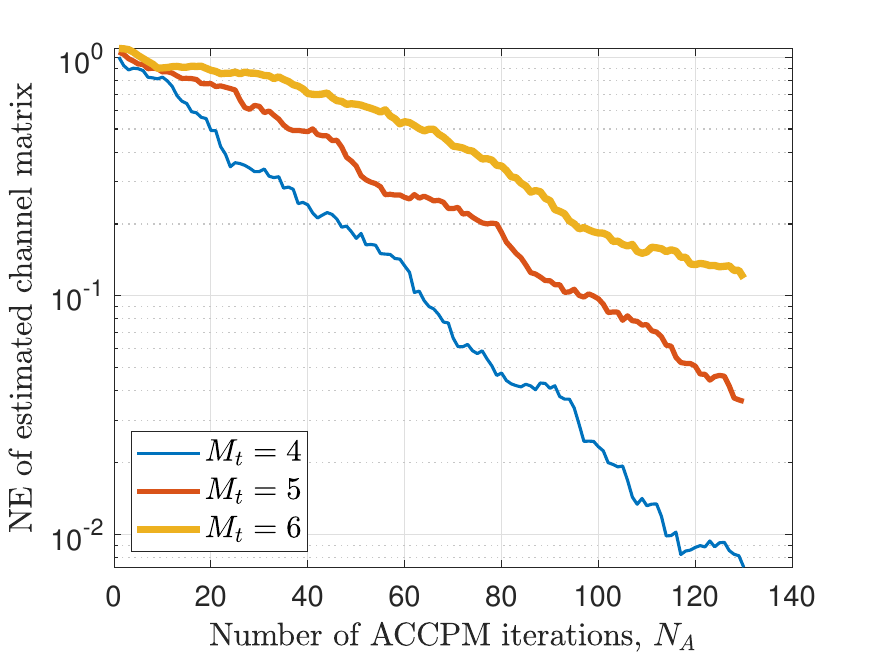}}
	\caption{NE of estimated channel matrix by applying the ACCPM versus iteration number $ N_A $ for different $ M_t $.}
	\label{fig_ACCPM_Convergence}
\end{figure}

In Fig. \ref{fig_ACCPM_Convergence}, we show the convergence behaviour of the ACCPM. We denote the number of iterations for the ACCPM as $ N_A $. Recall that the ACCPM is applied to learn a scaled version of the channel matrix $ {\bf{P}}\left( {{\bf{\bar v}}} \right) = \eta {\bf{\bar H}}_c^H{\bf{\bar v}}{{{\bf{\bar v}}}^H}{{{\bf{\bar H}}}_c} $.
To characterize the accuracy of the estimated scaled channel $ {{\bf{\tilde P}}^{{\rm{esti}}}} $, the normalized error (NE) of $ {{\bf{\tilde P}}^{{\rm{esti}}}} $ is defined as
\begin{align}
{\mathop{\rm NE}\nolimits} \big( {{{{\bf{\tilde P}}}^{{\rm{esti}}}}} \big) = \frac{{{{\big\| {{{{\bf{\tilde P}}}^{{\rm{esti,sc}}}} - {\bf{P}}} \big\|}_F}}}{{{{\left\| {\bf{P}} \right\|}_F}}},
\end{align}
where $ {{{\bf{\tilde P}}}^{{\rm{esti,sc}}}} = \frac{{{{\left\| {\bf{P}} \right\|}_F}}}{{{{\big\| {{{{\bf{\tilde P}}}^{{\rm{esti}}}}} \big\|}_F}}}{{{\bf{\tilde P}}}^{{\rm{esti}}}} $
is the scaled version of $ {{{\bf{\tilde P}}}^{{\rm{esti}}}} $ such that its Frobenius norm is the same as $ \bf{P} $.

It is observed that the ACCPM achieves an asymptotic decreasing estimation error with the number of feedback intervals, which shows its asymptotic convergence in practical implementation.
Fluctuations in the performance curves indicate that the NE of the estimated channel matrix is not strictly monotonically decreasing. It is reasonable since the cutting plane formed in each iteration has randomness.
We can also see that the convergence speed becomes slower as the number of ET's antennas increases. This is because the number of the elements in the channel matrix $ {\bf{P}} \in {\mathbb{C}^{{M_t} \times {M_t}}} $ to be estimated increases quadratically with $ M_t $, and a larger number of elements brings more difficulty of the channel matrix estimation.

\begin{figure*}[htbp]
	\centering
	\subfigure[$ M_t = 4 $]{
	\centering
	\includegraphics[width=0.31\textwidth]{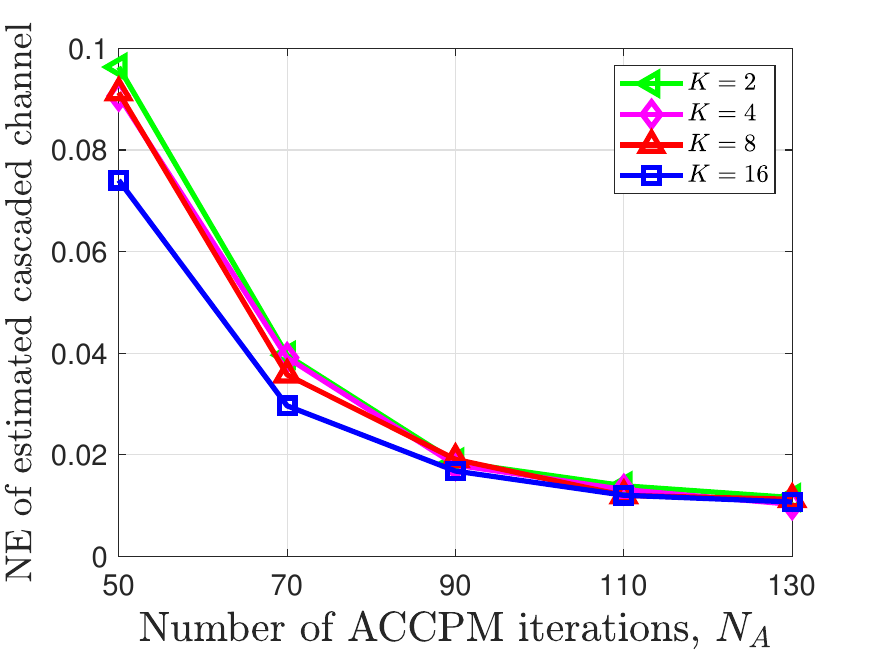}
	}
	\subfigure[$ M_t = 5 $]{
		\centering
		\includegraphics[width=0.31\textwidth]{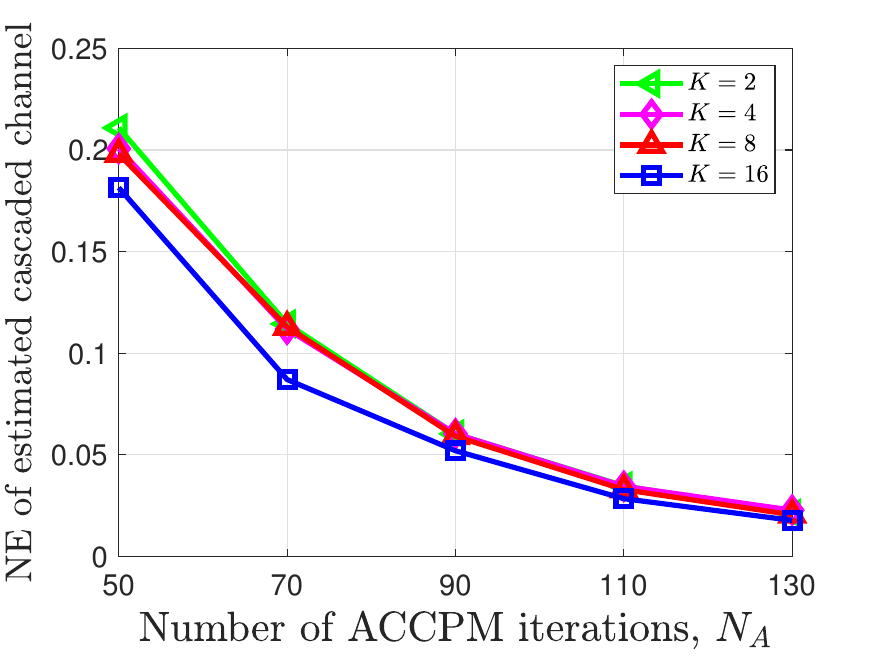}
	}
	\subfigure[$ M_t = 6 $]{
		\centering
		\includegraphics[width=0.31\textwidth]{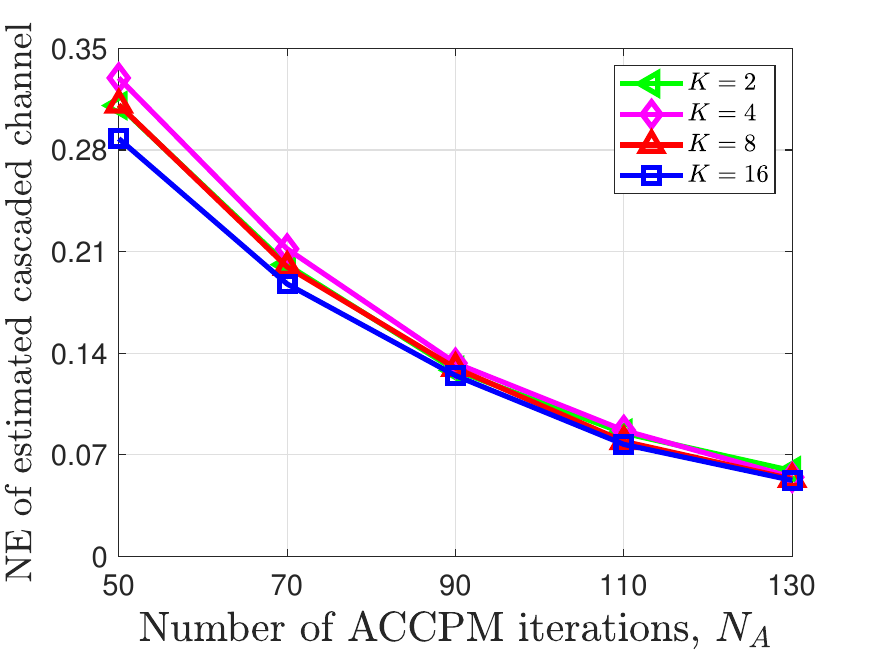}
	}
	\caption{Performance of the proposed one-bit feedback channel estimation method under different IRS group sizes.}
	\label{fig_OneBit_Feedback_NormError}
\end{figure*}

In Fig. \ref{fig_OneBit_Feedback_NormError}, we show the performance of our proposed one-bit-feedback-based channel estimation method. We plot the NE of the estimated cascaded channel matrix $ {\bf{\bar H}}_c^{{\rm{LS}}} $ versus the number of feedback intervals in ACCPM under different IRS group sizes. The NE is defined as 
\begin{align}
{\mathop{\rm NE}\nolimits} \left( {{\bf{\bar H}}_c^{{\rm{LS}}}} \right) = \arg \mathop {\min }\limits_\alpha  {\rm{ }}\frac{{{{\left\| {{e^{j\alpha }}{\bf{\bar H}}_c^{{\rm{LS,sc}}} - {{{\bf{\bar H}}}_c}} \right\|}_F}}}{{{{\left\| {{{{\bf{\bar H}}}_c}} \right\|}_F}}},
\end{align}
where $ {\bf{\bar H}}_c^{{\rm{LS,sc}}} = \frac{{{{\left\| {{{{\bf{\bar H}}}_c}} \right\|}_F}}}{{{{\left\| {{\bf{\bar H}}_c^{{\rm{LS}}}} \right\|}_F}}}{\bf{\bar H}}_c^{{\rm{LS}}} $.
For simplicity, we search for $ \alpha $ in $ \left[ {0,2\pi } \right) $ at $ {0.5^ \circ } $ intervals.

First, it is readily seen that a larger number of ACCPM iterations results in a more accurate estimated $ {{\bf{\bar H}}_c} $. This is due to the fact that more ACCPM iterations will result in smaller NE of $ \left\{ {{\bf{\tilde p}}\left( {{{{\bf{\bar v}}}_i}} \right)} \right\} $, whose linear combination, as shown in (\ref{Estimation_Hc}), constitutes the estimation of $ {{{\bf{\bar H}}}_c} $. 
Second, a smaller number of ET's antenna will accelerate the convergence of the ACCPM, as shown in Fig. \ref{fig_ACCPM_Convergence}. Thus, for the same $ K $ and $ N_A $, the NE of the estimated $ {{{\bf{\bar H}}}_c} $ increases with $ M_t $.
Finally, for fixed $ M_t $, the reduction in IRS group size tends to achieve a more rough estimate of the cascaded channel $ {{{\bf{\bar H}}}_c} $. This can be explained as the convergence rate of the ACCPM depends mainly on the number of ET's antenna, $ M_t $, and is not necessarily related to the IRS group size, $ K $. Therefore, a larger number of IRS groups leads to a larger cumulative error for the estimated $ {\bf{\bar H}}_c^{{\rm{LS}}} $ with a high probability from (\ref{Estimation_Hc}). 

\begin{figure*}[htbp]
	\centering
	\subfigure[$ M_t = 4 $]{
		\centering
		\includegraphics[width=0.31\textwidth]{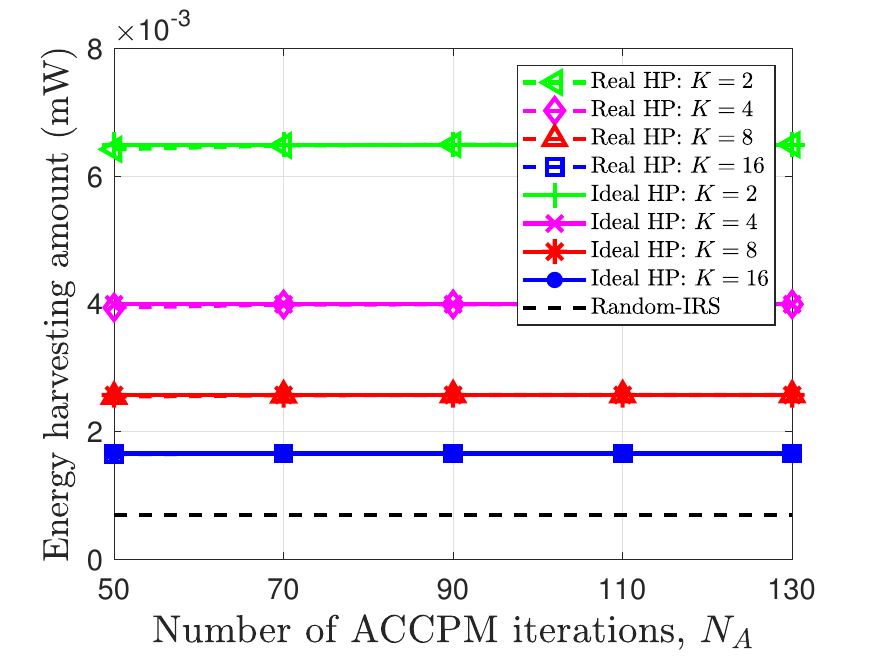}
	}
	\subfigure[$ M_t = 5 $]{
		\centering
		\includegraphics[width=0.31\textwidth]{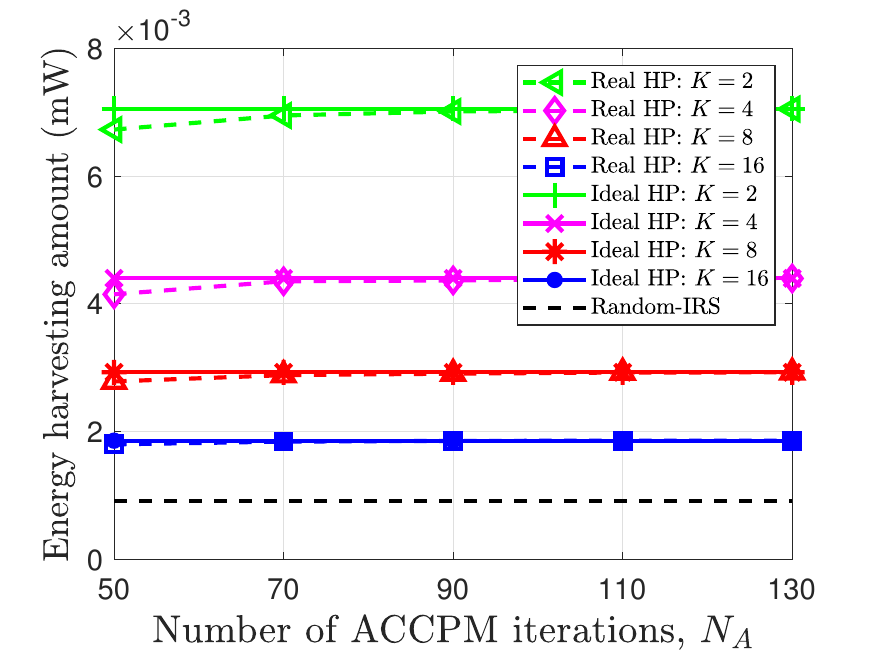}
	}
	\subfigure[$ M_t = 6 $]{
		\centering
		\includegraphics[width=0.31\textwidth]{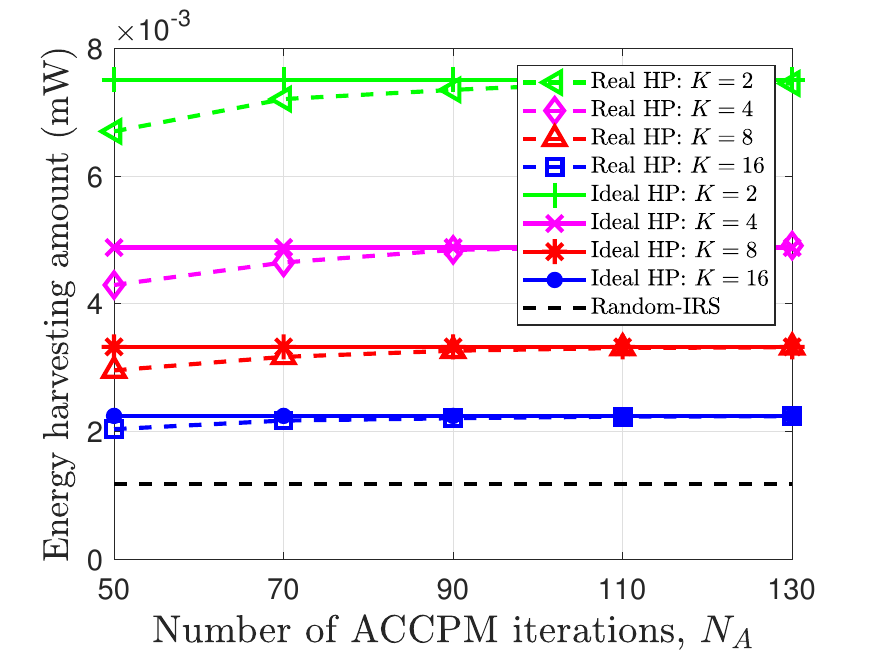}
	}
	\caption{The amount of harvested power at the ER by applying the proposed channel-estimation-based method for different IRS group sizes.}
	\label{fig_OneBit_Feedback_HarvestedPower}
\end{figure*}

In Fig. \ref{fig_OneBit_Feedback_HarvestedPower}, we plot the amount of the harvested energy at the ER versus the number of ACCPM iterations by applying the channel-estimation-based method under different IRS group sizes. We use the term ``Real HP'' to denote the harvested energy amount obtained by optimizing $ {\bf{Q}} $ and $ {{\bf{\bar v}}} $ based on the estimated cascaded channel $ {\bf{\bar H}}_c^{{\rm{LS}}} $. As a comparison, we plot the ideal harvested energy amount for different IRS group size, termed as ``Ideal HP'', where we assume that the cascaded channel $ {{{\bf{\bar H}}}_c} $ is perfectly estimated.
Combining with Fig. \ref{fig_OneBit_Feedback_NormError}, we can see that a more accurate channel estimation will reduce the gap between the real and ideal harvested energy amounts. In particular, when the NE of the estimated $ {{{\bf{\bar H}}}_c} $ is less than 0.1, the amount of energy harvested at the ER is almost the same as that of the perfect cascaded channel acquisition case. This means that the accuracy of the estimated channel is sufficient when its NE reaches below 0.1.
It is also seen that the reduction in IRS group size results in higher harvested power. This is because it helps a finer adjustment of the IRS reflection coefficients. Finally, the performance gain of our proposed channel-estimation-based method can be readily seen compared with the ``Random-IRS" case, where the IRS PSs are randomly selected, and the ET adopts the MRT precoding scheme.

\subsection{Distributed-Beamforming-Based Method}
\begin{figure}[htbp]
	\centerline{\includegraphics[width=0.4\textwidth]{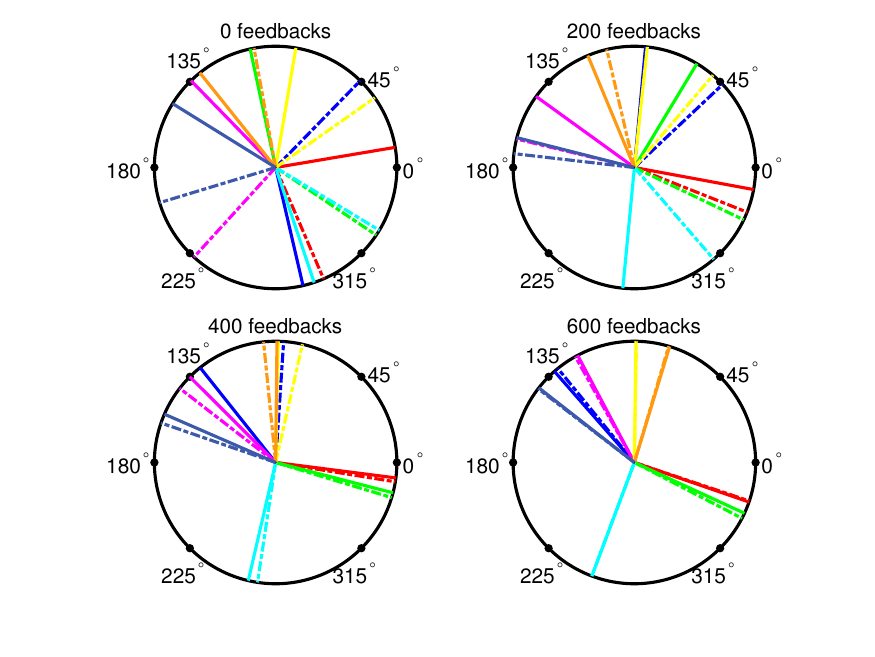}}
	\caption{Convergence behaviour of optimizing $ {\bm{\bar \theta}} $ based on the distributed beamforming method ($ M_t = 4 $, $ K = 8 $, and $ {{\bar \Delta }_0} = {2^ \circ } $).}
	\label{fig_Distri_Beamforming_Convergence_0}
\end{figure}

In Fig. \ref{fig_Distri_Beamforming_Convergence_0}, we show the convergence behaviour of the distributed beamforming algorithm to optimize $ {{\bm{\bar \theta }}} $ with $ M_t = 4 $, $ K = 8 $, and $ {{\bar \Delta }_0} = {2^ \circ } $. Each of the dash-dotted lines represent the current IRS group PS value, i.e., $ {{\bar \theta }_n}, \forall n \in \mathcal{J} $. While the solid lines represent the corresponding theoretical optimal values for each IRS group PS, respectively, i.e., $ \bar \theta _n^{{\rm{opti}}}\left[ n \right] = \arg \big( {\sum\limits_{j \ne n} {{\bf{\Phi }}\left( {n,j} \right){e^{j{{\bar \theta }_j}\left[ n \right]}}} } \big), \forall n \in \mathcal{J} $. 
It is readily seen that although each theoretically optimal PS value varies with the number of feedbacks, each updated IRS group PS gradually approaches its corresponding theoretical optimal value as the number of feedbacks increases, which responses to the statement in Proposition \ref{Theorem} that the distributed beamforming algorithm will eventually converge to a local maximum of problem (\ref{P4}) with probability 1.

\begin{figure}[htbp]
	\centerline{\includegraphics[width=0.4\textwidth]{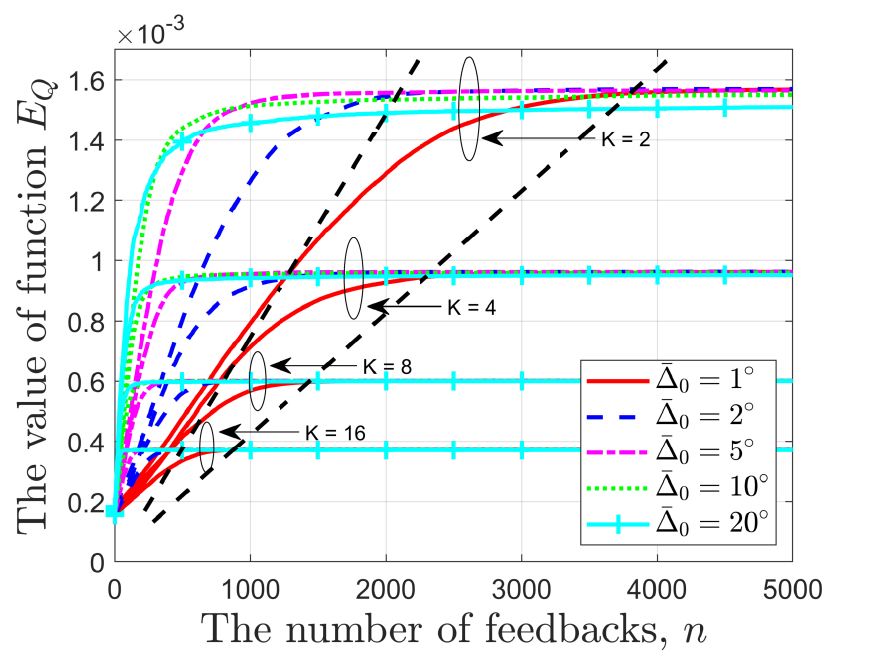}}
	\caption{Convergence behaviour of solving problem (\ref{P4}) by applying the distributed beamforming method for different values of $ K $ and $ {{\bar \Delta }_0} $ ($ M_t = 4 $).}
	\label{fig_Distri_Beamforming_Convergence_1}
\end{figure}

In Fig. \ref{fig_Distri_Beamforming_Convergence_1}, we show the convergence behavior of the distributed beamforming algorithm in solving problem (\ref{P4}), for different values of $ K $ and $ {\bar \Delta _0} $. First, we can see that for the same $ K $ and $ {\bar \Delta _0} $, a smaller IRS group size $ K $ leads to an increase in the number of IRS PSs that need to be optimized, and thus a larger number of feedbacks are required to converge. 
Further, from the two black dashed lines showing the relationship between the required number of feedbacks for convergence and $ K $ for $ {\bar \Delta _0} = {1^ \circ } $ and $ {\bar \Delta _0} = {2^ \circ } $, respectively, it can be observed that the number of feedbacks for the distributed beamforming algorithm is sublinear with respect to the number of IRS element groups $ J $, which demonstrates its excellent scalability for a large number of IRS PSs optimization. Second, for the same $ K $, a larger $ {\bar \Delta _0} $ results in a faster performance improvement for the distributed beamforming algorithm in the initial stage, as a larger range of phase perturbation makes it easier to find higher-quality IRS PS update results. However, it then leveled off at a lower performance, this is due to that the larger phase perturbation range leads to an increase in the proportion of invalid IRS phase shift exploration as the algorithm approaches convergence, and accordingly, the performance can hardly be improved. This inspires us that the value of $ {{{\bar \Delta }_0}} $ can be adapted dynamically during the optimization. For example, the value of $ {{{\bar \Delta }_0}} $ can be decreased linearly as the reduction of the update frequency of $ {{\bm{\bar \theta }}} $.

\begin{figure*}[htbp]
	\centering
	\subfigure[$ M_t = 4 $]{
		\centering
		\includegraphics[width=0.31\textwidth]{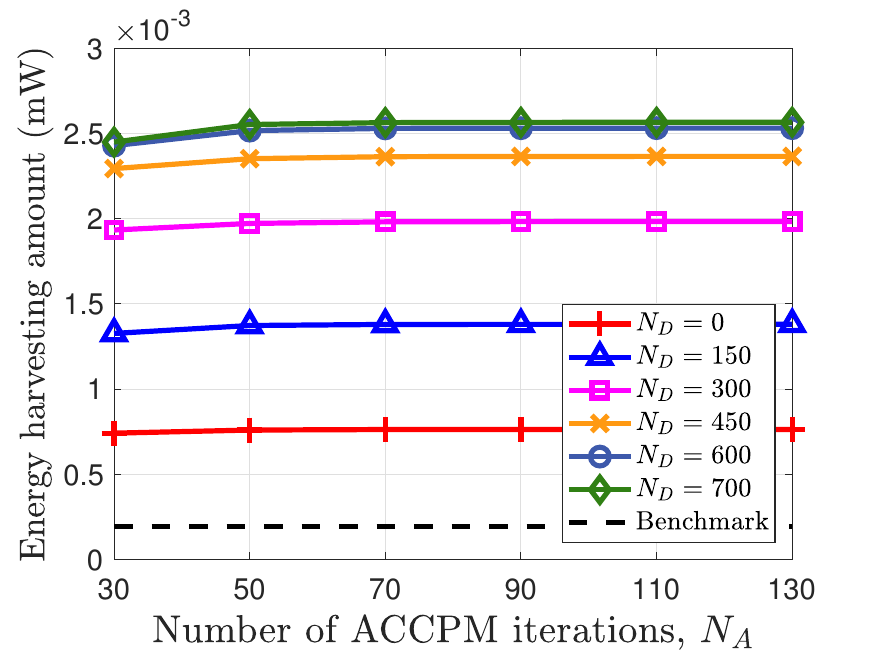}
	}
	\subfigure[$ M_t = 5 $]{
		\centering
		\includegraphics[width=0.31\textwidth]{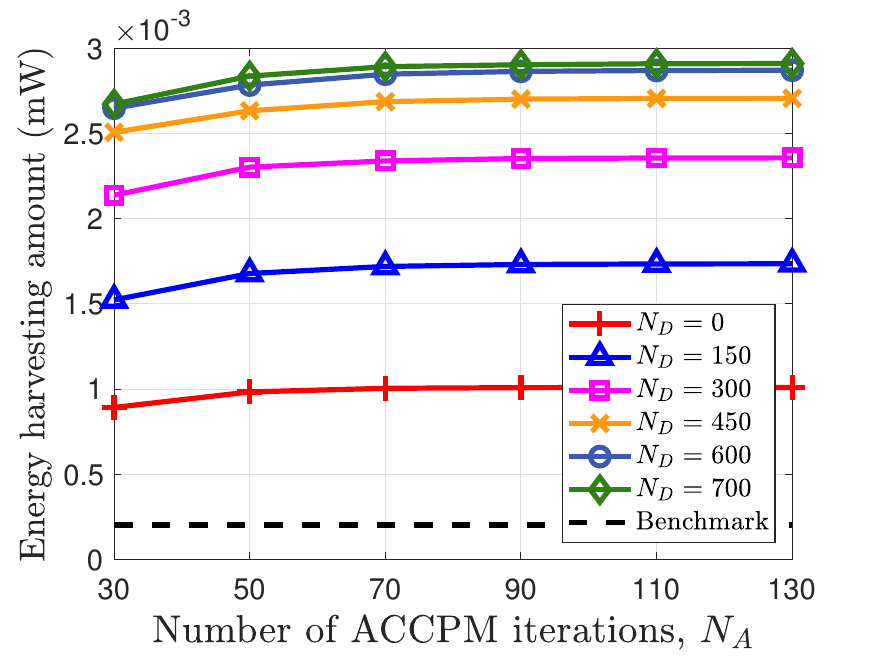}
	}
	\subfigure[$ M_t = 6 $]{
		\centering
		\includegraphics[width=0.31\textwidth]{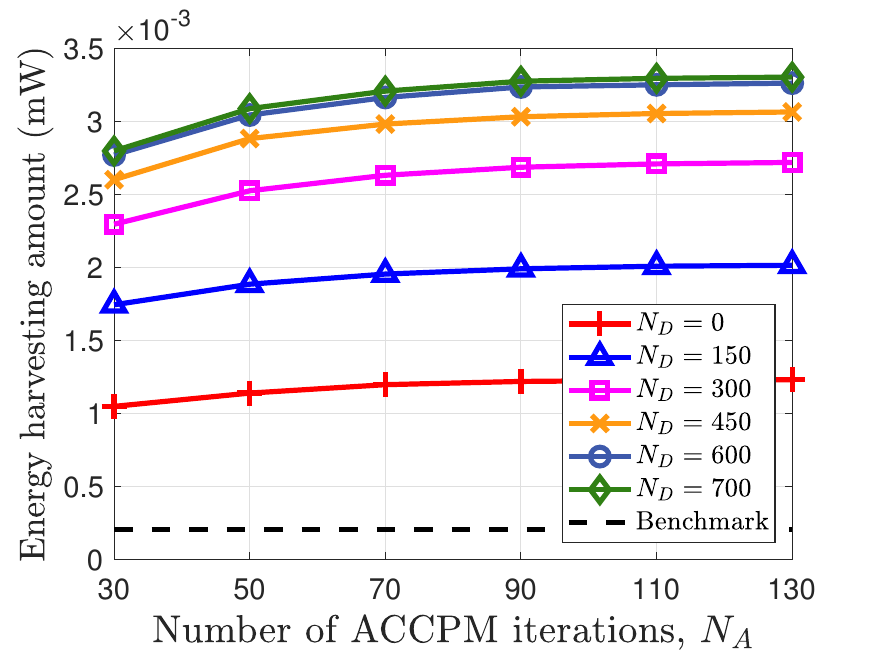}
	}
	\caption{Harvested energy amount by applying the distributed-beamforming-based method for different feedback numbers of ACCPM and distributed beamforming algorithm ($ K = 8 $, $ {\bar \Delta _0} = {2^ \circ } $).}
	\label{fig_DistriBased_Beamforming_Performance}
\end{figure*}

In Fig. \ref{fig_DistriBased_Beamforming_Performance}, we plot the ER's harvested energy amount obtained by applying the distributed-beamforming-based method for different feedback numbers of ACCPM (in the second stage) and distributed beamforming algorithm (in the first stage) with $ K = 8 $ and $ {{\bar \Delta }_0} = {2^ \circ } $. We use $ {N_D} $ to denote the feedback number for the distributed beamforming algorithm. Obviously, the amount of the harvested energy increases with the feedback numbers of both distributed beamforming procedure and ACCPM.
For all the $ M_t $ settings, the performance curves corresponding to $ {N_D} $ set to $ 600 $ and $ 700 $, respectively, are very close for all numbers of the ACCPM feedbacks, indicating that for the distributed beamforming algorithm, the performance is basically saturated at the current parameter setting ($ J = 8 $) when the number of feedbacks reaches $ 700 $.
Moreover, for a fixed $ {N_D} $, the number of ACCPM feedbacks required for convergence increases with $ M_t $, which is consistent with the convergence behaviour of the ACCPM shown in Fig. \ref{fig_ACCPM_Convergence}. 
In addition, the performance gain of the proposed distributed-beamforming-based method is verified by comparing with the benchmark scheme, where the IRS PSs are randomly selected and the transmit covariance matrix at the ET is set to $ {\bf{Q}} = \frac{{{P_b}}}{{{M_t}}}{{\bf{I}}_{{M_t}}} $.

\begin{figure}[H]
	\centerline{\includegraphics[width=0.38\textwidth]{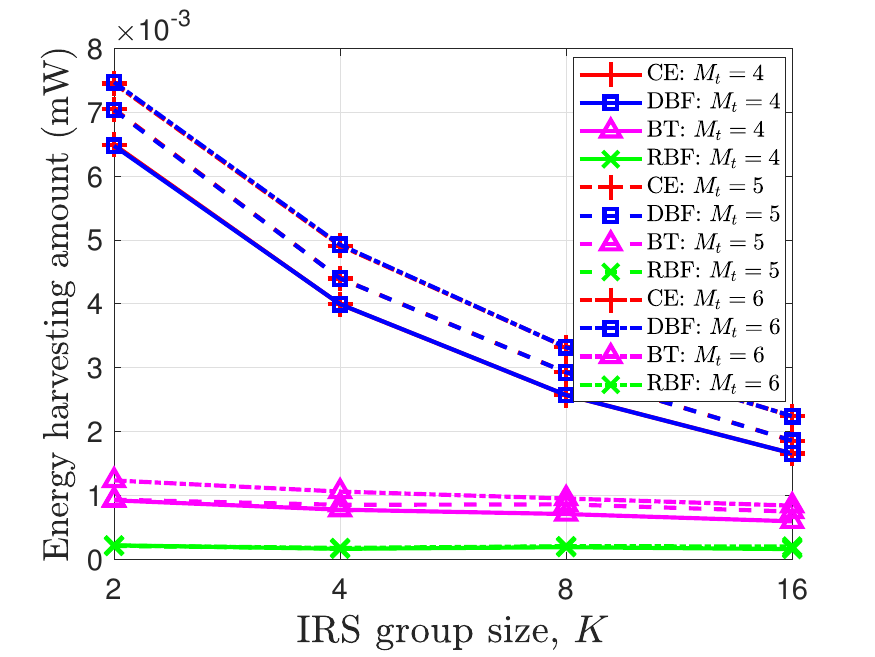}}
	\caption{The amount of harvested energy versus IRS group sizes for different $ M_t $ under different schemes.}
	\label{fig_Scheme_Comparison}
\end{figure}

In Fig. \ref{fig_Scheme_Comparison}, we compare the amount of the harvested energy obtained by our proposed joint beamforming design schemes with two benchmark schemes, namely, the beam training scheme and the random beamforming scheme. The simulation parameters are set as follows: 1) CE: for the channel-estimation-based method, the iteration number of ACCPM is set to $ N_A = 130 $; 2) DBF: for the distributed-beamforming-based method, we set the parameters of the first stage to $ N_D = 5000 $ and $ {\bar \Delta _0}{\rm{ = }}{2^ \circ } $, and the parameter of the second stage to $ N_A = 130 $; 3) BT: for the beam training scheme, the directional beams of the ET and IRS are selected from discrete Fourier Transform (DFT) codebook, and the entries of the transmit beamforming vector at the ET are set to phase-only complex variables with invariable amplitude \cite{9410435_Wang}. Moreover, exhaustive search is applied to find the optimal beam pair; 4) RBF: for the random beamforming scheme, we set $ {\bf{Q}} = \frac{{{P_b}}}{{{M_t}}}{\bf{I}} $, and the IRS PSs are randomly selected. From Fig. \ref{fig_Scheme_Comparison}, our proposed schemes achieve superior performance over the other schemes. Moreover, the channel-estimation-based scheme achieves almost the same performance as the distributed-beamforming-based scheme. The reason for the performance gap between the proposed schemes and the beam training scheme, especially for a smaller $ K $, is that Rayleigh fading is adopted as the small-scale fading for both ET-IRS and IRS-ER channels, which is not sparse, thus leading to a significant performance loss for the beam training scheme. The performance of the random beamforming scheme remains almost unchanged with $ M_t $ and $ K $, which demonstrates the importance of joint active and passive beamforming design for IRS-aided wireless system.

\section{Conclusion}
In this paper, we proposed two novel joint active and passive beamforming design schemes, namely, the channel-estimation-based method and the distributed-beamforming-based method, for an IRS-aided WET system exploiting only one-bit feedback from the ER to the ET.
Specifically, for the channel-estimation-based method, with the predetermined IRS PS matrix, we first estimated the scaled cascaded ET-IRS-ER channel by continually adjusting the transmit covariance matrix at the ET based on the ACCPM. Then, the joint beamforming design was performed based on the estimated cascaded channel by applying the existing optimization methods. 
While for the distributed-beamforming-based method, we first applied the distributed beamforming algorithm to optimize the IRS refection coefficients without knowing any involved CSI, which theoretically proven to converge to a local optimum almost surely. Then, for the given optimized IRS PSs, the ET's transmit covariance matrix was calculated according to the effective ET-ER channel learned by applying the ACCPM only once. 
Simulation results demonstrated the effectiveness of our proposed one-bit-feedback-based joint beamforming design schemes. More importantly, the proposed schemes were shown to be more appealing compared to existing pilot-based beamforming method and beam training method in terms of the ER's hardware complexity requirement and system performance, respectively. In particular, the proposed one-bit-feedback-based cascaded channel estimation method was also validated to have high accuracy. 

\appendices
\section{Proof of Proposition \ref{Proposition_Phase_difference}}\label{appendix_A}
The idea behind the proof is to resort to the derivative of $ {E_h}\left( \vartheta  \right) $ with respect to $ \vartheta  $. First, we have
\begin{align}
\frac{{\partial {E_h}}}{{\partial \vartheta }} =& 2\Re \left\{ {j\left( {{R_{i,j}}{e^{ - j\vartheta }}{\bf{\tilde p}}_i^H + {\bf{\tilde p}}_j^H} \right)} \right. \notag\\
&\big. {\left( {{R_{i,j}}{e^{j{\delta _{i,j}}}}{{{\bf{\tilde p}}}_i} + {{{\bf{\tilde p}}}_j}} \right)\big( {{e^{j\left( {\vartheta  - {\delta _{i,j}}} \right)}} - 1} \big)} \big\} \times \psi \left( \vartheta  \right),
\end{align}
where $ \psi \left( \vartheta  \right) = \frac{{R_{i,j}^2\big( {{{\left\| {{{{\bf{\tilde p}}}_i}} \right\|}^2}{{\left\| {{{{\bf{\tilde p}}}_j}} \right\|}^2} - {{\left| {{\bf{\tilde p}}_i^H{{{\bf{\tilde p}}}_j}} \right|}^2}} \big)}}{{{{\left| {{{\left( {{R_{i,j}}{e^{j\vartheta }}{{{\bf{\tilde p}}}_i} + {{{\bf{\tilde p}}}_j}} \right)}^H}\left( {{R_{i,j}}{e^{j\vartheta }}{{{\bf{\tilde p}}}_i} + {{{\bf{\tilde p}}}_j}} \right)} \right|}^2}}} $ is a positive real number, $ {{{\bf{\tilde p}}}_i} $ and $ {{{\bf{\tilde p}}}_j} $ are short for $ {\bf{\tilde p}}\left( {{{{\bf{\bar v}}}_{i}}} \right) $ and $ {\bf{\tilde p}}\left( {{{{\bf{\bar v}}}_{j}}} \right) $, respectively, and we define
$ {R_{i,j}} = \left| {{\gamma _i}} \right|/\left| {{\gamma _j}} \right| $, and $ {\delta _{i,j}} = \angle {\gamma _i} - \angle {\gamma _j} $. By setting the derivative of $ {E_h}\left( \vartheta  \right) $ w.r.t $ \vartheta  $ to zero, we have
\begin{align}
\frac{{\partial {E_h}}}{{\partial \vartheta }} = 0 & \Rightarrow   2\Re \left\{ {j\left( {{R_{i,j}}{e^{ - j\vartheta }}{\bf{\tilde p}}_i^H + {\bf{\tilde p}}_j^H} \right)} \right. \notag\\
&\big. {\left( {{R_{i,j}}{e^{j{\delta _{i,j}}}}{{{\bf{\tilde p}}}_i} + {{{\bf{\tilde p}}}_j}} \right)\big( {{e^{j\left( {\vartheta  - {\delta _{i,j}}} \right)}} - 1} \big)} \big\} = 0 \notag\\
& \Rightarrow f\left( \vartheta  \right) = 0,
\end{align} 
where
\begin{align}
f\left( \vartheta  \right) &= \big( {{{\left\| {{{{\bf{\tilde p}}}_j}} \right\|}^2} + R_{i,j}^2{{\left\| {{{{\bf{\tilde p}}}_i}} \right\|}^2}} \big)\sin \left( {\vartheta  - {\delta _{i,j}}} \right) \notag\\
&+ 2{R_{i,j}}\Re \left\{ {{\bf{\tilde p}}_i^H{{{\bf{\tilde p}}}_j}} \right\}\sin \vartheta  - 2{R_{i,j}}\sin {\delta _{i,j}}\Re \left\{ {{\bf{\tilde p}}_i^H{{{\bf{\tilde p}}}_j}} \right\} \notag\\
&= A\sin \big( {\vartheta  - \frac{{{\delta _{i,j}}}}{2} + \phi } \big) - {A_1}\sin {\delta _{i,j}},
\end{align}
with $ A = \sqrt {A_1^2 + A_2^2 + 2{A_1}{A_2}\cos \left( {{\delta _{i,j}}} \right)}  $, $ {A_1} = 2{R_{i,j}}\Re \left\{ {{\bf{\tilde p}}_i^H{{{\bf{\tilde p}}}_j}} \right\} $, $ {A_2} =  {{{\left\| {{{{\bf{\tilde p}}}_j}} \right\|}^2} + R_{i,j}^2{{\left\| {{{{\bf{\tilde p}}}_i}} \right\|}^2}}  $, and $ \phi  $ satisfies the equation $ \tan \phi  = \frac{{{A_1} - {A_2}}}{{{A_1} + {A_2}}}\tan \big( {\frac{{{\delta _{i,j}}}}{2}} \big) $. Mathematically, $ f\left( \vartheta  \right) = 0 $ is equivalent to the intersection of $ {f_1}\left( \vartheta  \right) = A\sin \big( {\vartheta  - \frac{{{\delta _{i,j}}}}{2} + \phi } \big) $ and $ {f_2} = {A_1}\sin {\delta _{i,j}} $. Since
\begin{subequations}\label{Intersection_proof}
\begin{align}
&{A^2} - A_1^2{\sin ^2}{\delta _{i,j}} \\
&= A_1^2{\cos ^2}{\delta _{i,j}} + A_2^2 + 2{A_1}{A_2}\cos \left( {{\delta _{i,j}}} \right)\\
&\mathop  \ge \limits^{\left( a \right)} A_1^2 + A_2^2 - 2\left| {{A_1}} \right|{A_2} > 0,
\end{align}
\end{subequations} 
where $ \left( a \right) $ is due to the fact that the minimum value of (\ref{Intersection_proof}b) is achieved when $ \cos \left( {{\delta _{i,j}}} \right) =  - \frac{{{A_1}}}{{\left| {{A_1}} \right|}} $, we have
\begin{align}
\left| A \right| > \left| {{A_1}\sin \left( {{\delta _{i,j}}} \right)} \right|.
\end{align}
Thus, as $ \vartheta  $ varies from $ 0 $ to $ 2\pi  $, $ {f_1}\left( \vartheta  \right) $ and $ {f_2} $ have two intersection points, i.e., there are two extreme points for $ {E_h}\left( \vartheta  \right) $ within the interval of $ \left[ {0,2\pi } \right) $, one is the local maximum point and the other is the local minimum point. Moreover, since $ {E_h}\left( \vartheta  \right) $ is a periodic function satisfying $ {E_h}\left( 0 \right) = {E_h}\left( {2\pi } \right) $, we can conclude that the local maximum is the global maximum. It is readily seen that the global maximum is obtained when $ \vartheta  = \angle {\gamma _i} - \angle {\gamma _j} $, thus the value of $ \angle {\gamma _i} - \angle {\gamma _j} $ can be obtained by finding the unique local maximum point of $ {E_h}\left( \vartheta  \right) $ within the interval of $ \left[ {0,2\pi } \right) $. This completes the proof.

\section{Proof of Proposition \ref{Proposition_IncreasePro}}\label{appendix_B}
Denote the $ n $th entry of $ {\bf{\bar v}} $ as $ {{\bar v}_n}, \forall n \in \mathcal{J} $. The function $ {E_Q}\left( {{\bf{\bar v}}} \right) $ in (\ref{P4}a) can be expanded as
\begin{align}\label{func_EQ}
{E_Q}\left( {{\bf{\bar v}}} \right) = \sum\limits_{i = 1}^J {{\bf{\Phi }}\left( {i,i} \right)}  + \sum\limits_{i = 1}^J {\sum\limits_{j = 1}^{i - 1} {2\Re \left\{ {{\bar v_i^ * }{\bf{\Phi }}\left( {i,j} \right){{\bar v}_j}} \right\}} }.
\end{align}
Then, the function of $ {E_Q}\left( {{\bf{\bar v}}} \right) $ with respect to the $ n $th entry of $ {{\bf{\bar v}}} $, i.e., $ {{{{\bar v}}}_n} $, can be written as
\begin{align}\label{func_EQ_vn}
{E_Q}\left( {{\bar v}_n} \right) = 2\Re \big\{ {{{\bar v}}_n^ * \sum\limits_{j \ne n} {{\bf{\Phi }}\left( {n,j} \right){{{{\bar v}}}_j}} } \big\} + {\text{constant}}.
\end{align}
Applying the chain rule, the derivative of the function $ {E_Q}\left( {{\bf{\bar v}}\left( {{\bm{\bar \theta }}} \right)} \right) $ with respect to $ {{{{\bar \theta }}}_n} $ is given by
\begin{align}\label{derivative_EQ}
\frac{{\partial {E_Q}\left( {{\bf{\bar v}}\left( {{\bm{\bar \theta }}} \right)} \right)}}{{\partial {{\bar \theta }_n}}} &= \frac{{\partial {E_Q}\left( {{\bf{\bar v}}} \right)}}{{\partial {{\bar v}_n}}}\frac{{\partial {{\bar v}_n}}}{{\partial {{\bar \theta }_n}}} + \frac{{\partial {E_Q}\left( {{\bf{\bar v}}} \right)}}{{\partial \bar v_n^ * }}\frac{{\partial \bar v_n^ * }}{{\partial {{\bar \theta }_n}}} \notag\\
&= 2\Im \big\{ {{e^{ - j{{\bar \theta }_n}}}\big( {\sum\limits_{j \ne n} {{\bf{\Phi }}\left( {n,j} \right){{\bar v}_j}} } \big)} \big\}.
\end{align}

For the second term on the right hand side of (\ref{func_EQ}), we have
\begin{align}
\sum\limits_{i = 1}^J {\sum\limits_{j = 1}^{i - 1} {2\Re \left\{ {\bar v_i^ * {\bf{\Phi }}\left( {i,j} \right){{\bar v}_j}} \right\}} } = 2\sum\limits_{i = 1}^J {\big| {\sum\limits_{j = 1}^{i - 1} {{\bf{\Phi }}\left( {i,j} \right){{\bar v}_j}} } \big|\cos \left( {{\varphi _i}} \right)} ,
\end{align}
where $ {\varphi _i} = \arg \big( {\sum\limits_{j = 1}^{i - 1} {{\bf{\Phi }}\left( {i,j} \right){{\bar v}_j}} } \big) - \bar \theta_i  $ is the angle difference between the current $ {{\bar \theta }_i} $ and the local maximum. For simplicity, we assume that $ \left\{ {\left| {{\varphi _i}} \right|} \right\}_{i = 1}^J $ are sorted such that $ \left| {{\varphi _1}} \right| \ge \left| {{\varphi _2}} \right| \ge  \cdots  \ge \left| {{\varphi _J}} \right| $. Thus we have
\begin{align}
\sum\limits_{i = 1}^J {\sum\limits_{j = 1}^{i - 1} {2\Re \left\{ {\bar v_i^ * {\bf{\Phi }}\left( {i,j} \right){{\bar v}_j}} \right\}} }  \ge 2\cos \left( {{\varphi _1}} \right)\sum\limits_{i = 1}^J {\big| {\sum\limits_{j = 1}^{i - 1} {{\bf{\Phi }}\left( {i,j} \right){{\bar v}_j}} } \big|}.
\end{align}
Assuming that $ {E_Q}\left( {{\bf{\bar v}}\left( {{\bm{\bar \theta }}} \right)} \right) \le {Y^{{\rm{opt}}}} - \epsilon $, from (\ref{derivative_EQ}) we conclude that $ \left| {{\varphi _1}} \right| > 0 $. Further, we have
\begin{align}
&\sum\limits_{i = 1}^J {{\bf{\Phi }}\left( {i,i} \right)}  + 2\cos \left( {{\varphi _1}} \right)\sum\limits_{i = 1}^J {\big| {\sum\limits_{j = 1}^{i - 1} {{\bf{\Phi }}\left( {i,j} \right){{\bar v}_j}} } \big|}  \le {Y^{{\rm{opt}}}} - \epsilon \notag\\
&\Rightarrow \left| {{\varphi _1}} \right| \ge {\phi _\epsilon} \buildrel \Delta \over = \arccos \left( {\frac{{{Y^{{\rm{opt}}}} - \epsilon - {\sum\limits_i {{\bf{\Phi }}\left( {i,i} \right)} } }}{{2\sum\limits_{i = 1}^J {\big| {\sum\limits_{j = 1}^{i - 1} {{\bf{\Phi }}\left( {i,j} \right){{\bar v}_j}} } \big|} }}} \right).
\end{align}

For ease of analysis, we treat the perturbation $ \bm{\delta} $ applied to $ {{\bm{\bar \theta }}} $ as each perturbation $ \delta_i $ applied to $ {{\bar \theta }_i} $ one by one. Now, we choose a phase perturbation $ \delta_1 $ such that $ \left| {{\varphi _1}} \right| $ is decreased. This makes the amount of harvested energy at the ER increase. Without loss of generality, we assume $ {\varphi _1} > 0 $, then we need to choose a $ \delta_1 > 0 $. Consider $ {\delta _1} \in \left( {\frac{{{\Delta _0}}}{2},{\Delta _0}} \right) $, we have
\begin{align}
{\rho _1} \buildrel \Delta \over = \Pr \big( {\frac{{{\Delta _0}}}{2} \le {\delta _1} \le {\Delta _0}} \big) = \int_{\frac{{{\Delta _0}}}{2}}^{{\Delta _0}} {g\left( x \right)dx}  > 0,
\end{align}
and from (\ref{func_EQ_vn}), we obtain
\begin{align}
&{E_Q}\left( {{{\bar v}_1}\left( {{{\bar \theta }_1} + {\delta _1}} \right)} \right) - {E_Q}\left( {{{\bar v}_1}\left( {{{\bar \theta }_1}} \right)} \right) \notag\\
&= 2\Re \big\{ {\big( {{e^{ - j\left( {{{\bar \theta }_1} + {\delta _1}} \right)}} - {e^{ - j{{\bar \theta }_1}}}} \big)\sum\limits_{j \ne 1} {{\bf{\Phi }}\left( {n,j} \right){{\bar v}_j}} } \big\} \notag\\
&= 2\big| {\sum\limits_{j \ne 1} {{\bf{\Phi }}\left( {n,j} \right){{\bar v}_j}} } \big|\left( {\cos \left( {{\varphi _1} - {\delta _1}} \right) - \cos \left( {{\varphi _1}} \right)} \right) \notag\\
&> 2\big| {\sum\limits_{j \ne 1} {{\bf{\Phi }}\left( {n,j} \right){{\bar v}_j}} } \big|\frac{{{\Delta _0}}}{2}\sin \big( {{\phi _\epsilon} - \frac{{{\Delta _0}}}{2}} \big) = 2{\epsilon_1},
\end{align}
where $ {\epsilon_1} \buildrel \Delta \over = \frac{{{\Delta _0}}}{2}\big| {\sum\limits_{j \ne 1} {{\bf{\Phi }}\left( {n,j} \right){{\bar v}_j}} } \big|\sin \left( {{\phi _\epsilon} - \frac{{{\Delta _0}}}{2}} \right) $.
To obtain a positive value of $ {E_Q}\left( {{\bf{\bar v}}\left( {{\bm{\bar \theta }} + {\bm{\delta }}} \right)} \right) - {E_Q}\left( {{\bf{\bar v}}\left( {{\bm{\bar \theta }}} \right)} \right) $, a safe measure is to ensure that the other IRS phase shift perturbations $ {\left\{ {{\delta _j}} \right\}_{j \ne 1}} $ are small enough compared to $ \delta_1 $. From (\ref{func_EQ}), we have (\ref{EQ_improvement}), which is shown at the top of the next page. Since $ {E_Q}\left( {{\bf{\bar v}}\left( {{\bm{\bar \theta }}} \right)} \right) $ is continuous in each of the phases $ {{\bar \theta }_i} $, we can always find a $ \epsilon_1 > 0, \forall i > 1 $, to satisfy (\ref{eq_ige1}), which is also shown at the top of the next page. Due to
\begin{figure*}[!t]
	\normalsize
	\begin{align}\label{EQ_improvement}
	&{E_Q}\left( {{\bf{\bar v}}\left( {{\bm{\bar \theta }} + {\bm{\delta }}} \right)} \right) - {E_Q}\left( {{\bf{\bar v}}\left( {{\bm{\bar \theta }}} \right)} \right) \notag\\
	&= {E_Q}\left( {{{\bar v}_1}\left( {{{\bar \theta }_1} + {\delta _1}} \right)} \right) - {E_Q}\left( {{{\bar v}_1}\left( {{{\bar \theta }_1}} \right)} \right)  + \sum\limits_{i > 1} {{E_Q}\left( {{{\bar v}_i}\left( {{{\bar \theta }_i} + {\delta _i}} \right)|{{\bar \theta }_j} \leftarrow {{\bar \theta }_j} + {\delta _j},\forall j < i} \right) - {E_Q}\left( {{{\bar v}_i}\left( {{{\bar \theta }_i}} \right)|{{\bar \theta }_j} \leftarrow {{\bar \theta }_j} + {\delta _j},\forall j < i} \right)} \notag\\
	&> 2{\epsilon_1} + 2\Re \Big\{ {\sum\limits_{i > 1} {{e^{ - j{{\bar \theta }_i}}}\left( {{e^{ - j{\delta _i}}} - 1} \right)\Big( {\sum\limits_{j < i} {{\bf{\Phi }}\left( {i,j} \right){e^{j\left( {{{\bar \theta }_j} + {\delta _j}} \right)}}}  + \sum\limits_{j > i} {{\bf{\Phi }}\left( {i,j} \right){e^{j{{\bar \theta }_j}}}} } \Big)} } \Big\}.
	\end{align}	
	\hrulefill
	
	\begin{align}\label{eq_ige1}
	\bigg| {2\Re \Big\{ {{e^{ - j{{\bar \theta }_i}}}\left( {{e^{ - j{\delta _i}}} - 1} \right)\Big( {\sum\limits_{j < i} {{\bf{\Phi }}\left( {i,j} \right){e^{j\left( {{{\bar \theta }_j} + {\delta _j}} \right)}}}  + \sum\limits_{j > i} {{\bf{\Phi }}\left( {i,j} \right){e^{j{{\bar \theta }_j}}}} } \Big)} \Big\}} \bigg| < \frac{{{\epsilon_1}}}{{J - 1}},\forall \big| {{\delta _i}} \big| < {\epsilon_i},i > 1.
	\end{align}
	\hrulefill
\end{figure*}
\begin{align}
&\left( {{\text{The left hand side of }}\left( \text{\ref{eq_ige1}} \right)} \right) \notag\\
&\le 2\big| {{e^{ - j{{\bar \theta }_i}}}\left( {{e^{ - j{\delta _i}}} - 1} \right)} \big| \notag\\
& \qquad \times \Big| {\Big( {\sum\limits_{j < i} {{\bf{\Phi }}\left( {i,j} \right){e^{j\left( {{{\bar \theta }_j} + {\delta _j}} \right)}}}  + \sum\limits_{j > i} {{\bf{\Phi }}\left( {i,j} \right){e^{j{{\bar \theta }_j}}}} } \Big)} \Big| \notag\\
&\le 4\big| {\sin \big( {\frac{{{\delta _i}}}{2}} \big)} \big|\sum\limits_{j \ne i} {\left| {{\bf{\Phi }}\left( {i,j} \right)} \right|},   
\end{align}
it can be readily seen that $ {\epsilon_i} = 2\arcsin \big( {\frac{{{\epsilon_1}}}{{4\left( {J - 1} \right)\sum\nolimits_{j \ne i} {\left| {{\bf{\Phi }}\left( {i,j} \right)} \right|} }}} \big) $ satisfies (\ref{eq_ige1}). With $ {\left\{ {{\delta _i}} \right\}_{i > 1}} $ chosen that satisfies (\ref{eq_ige1}), we have
\begin{align}\label{Conclusion_appen_B}
{E_Q}\left( {{\bf{\bar v}}\left( {{\bm{\bar \theta }} + {\bm{\delta }}} \right)} \right) - {E_Q}\left( {{\bf{\bar v}}\left( {{\bm{\bar \theta }}} \right)} \right) \ge {\epsilon_1}.
\end{align}
Moreover, since the PDF $ g\left( {{\delta _i}} \right) $ is bounded away from zero in each of the intervals $ \left( { - {\epsilon_i},{\epsilon_i}} \right) $, the probability $ \rho_i $ of choosing $ \delta_i $ to satisfies (\ref{eq_ige1}) is nonzero, i.e., $ \rho_i > 0 $. Finally, we can see that all the entries in $ \left\{ {{\delta _i}} \right\}_{i = 1}^J $ are chosen independently, thus (\ref{Conclusion_appen_B}) holds at least with probability $ \rho  = \prod\nolimits_i {{\rho _i}}  > 0 $. This thus completes the proof.

\bibliographystyle{IEEEtran}
\bibliography{mybibfile}

\end{document}